\documentclass[oldversion]{aa}
\usepackage{txfonts}
\usepackage{graphicx}
\usepackage{dcolumn}
\usepackage{subfigure}
\usepackage{natbib}
\bibpunct{(}{)}{;}{a}{}{,}

\newcommand{\rst}{\rho^*}

\newcommand{\pa}[1]{\partial_#1}
\newcommand{\pau}[1]{\partial^#1}
\newcommand{\hubf}{\mathbf{\hat{u}}}
\newcommand{\hu}{\hat{u}}

\begin{document}

   \title{Relativistic neutron star merger simulations with non-zero
	temperature equations of state\\
	I. Variation of binary parameters and equation of state}
   \titlerunning{Relativistic NSM simulations with non-zero
   temperature EoSs}

   \author{R. Oechslin \and H.-T. Janka \and A. Marek}
\institute{Max-Planck-Institut f\"ur Astrophysik, Karl-Schwarzschild-Str. 1, D-85741 Garching, Germany}
\offprints{R. Oechslin}
\date{Received / accepted} 

\abstract{
        An extended set of binary neutron star (NS) merger
        simulations is performed with an approximative treatment of
        general relativity to systematically investigate the influence 
	of the nuclear equation of state (EoS), the neutron star masses, and
        the NS spin states prior to merging.\\
        The general relativistic hydrodynamics simulations are
        based on a conformally flat approximation to the Einstein
        equations and a Smoothed Particle Hydrodynamics code for the
        gas treatment. We employ the two non-zero temperature EoSs of
        Shen et al. (1998a,b) and Lattimer \& Swesty (1991), which
        represent a ``harder'' and a ``softer'' behavior,
        respectively, with characteristic differences in the
        incompressibility at supernuclear densities
        and in the maximum mass of nonrotating, cold neutron stars. In
        addition, we use the cold EoS of Akmal et al. (1998) with
        a simple ideal-gas-like extension according to
        Shibata \& Taniguchi (2006), in order to compare with their results, 
	and an ideal-gas EoS with parameters fitted to the
        supernuclear part of the Shen-EoS.
	We estimate the mass sitting in a dilute ``torus'' around the
        future black hole (BH)
        by requiring the specific angular momentum of the torus matter
        to be larger than the angular momentum of the ISCO around a
        Kerr BH with the mass and spin parameter of the compact
        central remnant. The dynamics and outcome of the models is found to depend
        strongly on the EoS and on the binary parameters.  
	Larger torus masses are found for asymmetric systems (up to
        $\sim 0.3M_\odot$ for a mass ratio of 0.55), for large initial NSs,
        and for a NS spin state which corresponds to a larger
	total angular
        momentum. We find that the postmerger remnant collapses either 
        immediately or after a short time when employing the soft EoS
        of Lattimer\& Swesty, whereas no sign of post-merging collapse
        is found within tens of dynamical timescales for
        all other EoSs used. The typical temperatures in the torus are
        found to be about $3-10$ MeV depending on the strength of the shear motion
        at the collision interface between the NSs and thus depending
        on the initial NS spins.
	About $10^{-3}-10^{-2} M_\odot$ of NS matter become
        gravitationally unbound during or right after the merging
        process. This matter 
        consists of a hot/high-entropy component from the collision
        interface and (only in case of asymmetric systems) of a
        cool/low-entropy component from the spiral arm tips.  }


   \maketitle

\section{Introduction}

Merger events of double neutron star (DNS) binaries do not only belong to the
strongest known sources of gravitational wave (GW) radiation, they are
also widely favored as origin for the class of short, hard
GRBs \citep{blinnikov1984, paczynski1986, eichler1989,
paczynski1991, narayan1992, oechslin2006a}. This possible link seems
to be supported by recent short GRBs with determined redshift \citep[see e.g.][]{gehrels2005nature, fox2005, berger2005}. Theoretical gravitational waveforms from
simulations are needed in the future detection process as templates
for filtering out the actual signal from the noisy detector output. The gravitational wave
signal of the late inspiral and merging phases contains information on the
nuclear EoS and on the binary parameters such as NS masses and
spins.

Knowledge about the merging is also needed as a basic ingredient
for theoretical short GRB models \citep{aloy2005}. The post-merging
black hole (BH)-torus configuration serves as initial data for such
models and thus, information about the torus parameters is essential in this context.

Prior to coalescence, the DNS binary spirals in due to energy and
angular momentum loss by gravitational radiation. This inspiral
phase lasts millions of years and can be very accurately modelled in
the post-Newtonian framework \citep[e.g.][]{damour2001,damour2002}. In
contrast, the coalescence itself happens on a dynamical timescale of milliseconds and simulating such a process
naturally involves three-dimensional general relativistic
hydrodynamics. In addition, different aspects of physics enter the
problem at some stage. Due to the high compactness, gravitational
interaction has to be described by full general relativity (GR). Nuclear and
particle physics enter in the description of the hot and dense
neutron star fluid via an equation of state (EoS) and in the treatment
of the neutrino generation after the actual merging event. Finally,
the possible formation of heavy elements is of interest when matter
gets ejected during the coalescence.

Up to now, attempts to investigate the process have concentrated
either on the relativistic aspects while greatly simplifying the
microphysics, or a nuclear physics based EoS
was used together with an approximative neutrino transport while
describing gravity in a Newtonian framework.
The field was pioneered by the work of \citet{oohara1989}. Relativistic
aspects have been considered in \citet{oohara1992,ayal2001,faber2000}
and \citet{faber2002} in the post-Newtonian (PN) framework, which
however breaks down when gravity becomes too strong. The
conformal flatness approach, a middle ground between PN and full
general relativity (GR), has been chosen in \citet{wilson1996,
oechslin2002, oechslin2004, faber2004}, and \citet{oechslin2006a}. A fully relativistic treatment has
been adopted by \citet{shibata2002, shibata2005}, and
\citet{shibata2006}.

On the other hand, Newtonian simulations with microphysical
improvements have been achieved in
\citet{ruffert1996,rosswog1999,ruffert2001}, and \citet{rosswog2003}. These
works implemented either the nuclear-physics based,
non-zero-temperature equation of state (EoS) of
\citet{lattimer1991} or the one of \citet{shen1998,shen1998b}. Furthermore,
neutrino physics has been implemented using a neutrino-trapping
scheme.

First steps to unite these two ``branches'' have been taken on the one hand by \citet{oechslin2004}, who considered within the conformal flatness
approximation to GR the influence of different microphysical
zero-temperature EoSs on the inspiral and merger dynamics, and, on the
other hand by
\citet{shibata2005} and \citet{shibata2006}, who implemented besides full GR an analytic fit
to several physical zero-temperature
EoSs together with an ideal-gas-like extension that approximatively
accounts for finite-temperature effects.

Using an extended set of hydrodynamics simulations, the present work
studies the dependence of the merging and the subsequent postmerger evolution
on the nuclear EoS and on the binary parameters like the NS masses and
the initial NS spins. 
Similar aims, in Newtonian and post-Newtonian gravity, but without
such combination of GR and detailed microphysics have been
pursued by several authors \citep{rosswog2000, faber2001, faber2002},
mainly focusing on the gravitational wave signal and the amount of
gravitationally unbound matter.

Our paper is organized as follows. In Sect. \ref{sect:formalism} and
in Appendix \ref{app:hydro}, we sketch the formalism on
which our simulations are based and its implementation in our
numerical scheme.  In Sect. \ref{sect:init}, the
different initial models for our simulations and their construction is
described. The results are presented in
Sect. \ref{sect:results}. Finally, our conclusions are summarized in Sect. \ref{sect:conclusions}.

\section{The Formalism}
\label{sect:formalism}
\subsection{Hydrodynamics}
We employ an improved version of our relativistic smoothed particle
hydrodynamics (SPH) code \citep{oechslin2002,oechslin2004}, which solves the
relativistic hydrodynamics equations together with the Einstein field
equations in the conformally flat approximation (CFC; \citealt{Isenberg1980},
\citealt{wilson1996}). The detailed formalism and its implementation
are sketched in Appendix \ref{app:hydro}.

The improved code version now includes two finite-temperature nuclear
EoSs, namely the ones of \citet{shen1998,shen1998b} and of
\citet{lattimer1991}. They allow for a microphysically-based treatment
of matter from low densities to well above nuclear saturation density. Their
non-zero-temperature character also allows for an investigation of the
thermal evolution in regions where shock heating is present. We have also implemented the new SPH artificial viscosity (AV)
formalism of \citet{chow1997}. It replaces the ``standard'' formalism
\citep{monaghan1992} and the one proposed by \citet{siegler2000}, which were used in
our old code version. The AV interaction is determined in \citet{chow1997} by
approximately solving a Riemann problem between particle neighbours.

The revised backreaction formalism mimicking the energy and angular momentum
dissipation by gravitational waves now follows the lines of the
formalism of \citet[][BDS]{BDS}, but includes some higher-order
terms as given by \cite{faye2003} (see Appendix \ref{app:hydro}). The reactive contribution to the metric
coefficients is added to the CFC-approximated metric terms.

Neutrino physics is not implemented as we consider the backreaction of
neutrino emission on the fluid as negligeable on timescales on the
order of $\sim 10$ms considered here. As a consequence, the electron
fraction $Y_e$, once determined from the initial conditions, does not
change within a fluid element and is simply advected with the moving fluid. 

\subsection{The Equations of State}
\label{sect:eos}
We consider a set of four different EoSs:

\begin{itemize}
\item The EoS of \citet[][``Shen-EoS'']{shen1998,shen1998b}, 
\item the EoS of \citet[][``LS-EoS'']{lattimer1991},
\item the EoS of \citet[][``APR-EoS'']{akmal1998} extended with an ideal
gas-like thermal contribution, and
\item an ideal-gas EoS with adiabatic index and adiabatic constant
fitted to the supernuclear part of the Shen-EoS.
\end{itemize}

The Shen-EoS uses a phenomenological relativistic mean field (RMF) approach which aims at
reproducing results from Relativistic Br\"uckner-Hartree-Fock (RBHF)
calculations. It has a relatively high incompressibility modulus of
$K_0=280$MeV and thus leads to large NS-radii and maximum masses
(see Fig. \ref{fig:mrprofiles}). We implement the EoS in tabulated form
with $\rho$ (baryonic density), $T$ (temperature), and $Y_e$ (electron
fraction) as input quantities. Since our hydro code evolves the
specific internal energy $\epsilon$ rather than the temperature (via the energy equation (\ref{eqn:energyeqn2}) and the
relation (\ref{eqn:eevol}) in Appendix \ref{app:hydro}), the EoS has to be evaluated
implicitly for given $\rho$,$Y_e$, and $\epsilon$.

The LS-EoS employs a finite-temperature compressible liquid
droplet model for nuclei \citep{LLPR}. Our version of the LS-EoS
has been constructed using an incompressibility modulus of $K_0=180$MeV, thus
generally providing lower pressure values than the Shen-EoS for given
$\rho, T$, and $Y_e$. Consequently, the NS radii are much smaller and the maximum
NS mass allowed by the LS-EoS is marginally compatible with the NS
mass of $2.1\pm0.2 M_\odot$ recently measured by
\citet{nice2005}. The numerical implementation of the LS-EoS in our
code is similar to the one of the Shen-EoS.

\begin{figure}
\begin{center}
\includegraphics[width=8cm]{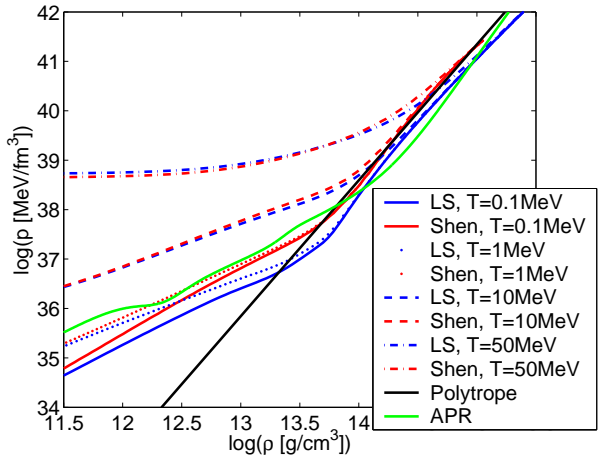}
\caption{Pressure versus density for a set of representative
temperatures for the two considered physical EoSs of
\citet{lattimer1991} and of \citet{shen1998,shen1998b}. The electron
fraction is set to $Y_e=0.05$. For comparison, we also show the
polytropic relation $p=K\rho^\Gamma$ used in this work with $K=30000,
\Gamma=2.75$ and the APR-EoS for $T=0$.}
\label{fig:p-rho-plot}
\end{center}
\end{figure}

\begin{figure}
\begin{center}
(a)
\includegraphics[width=7.5cm]{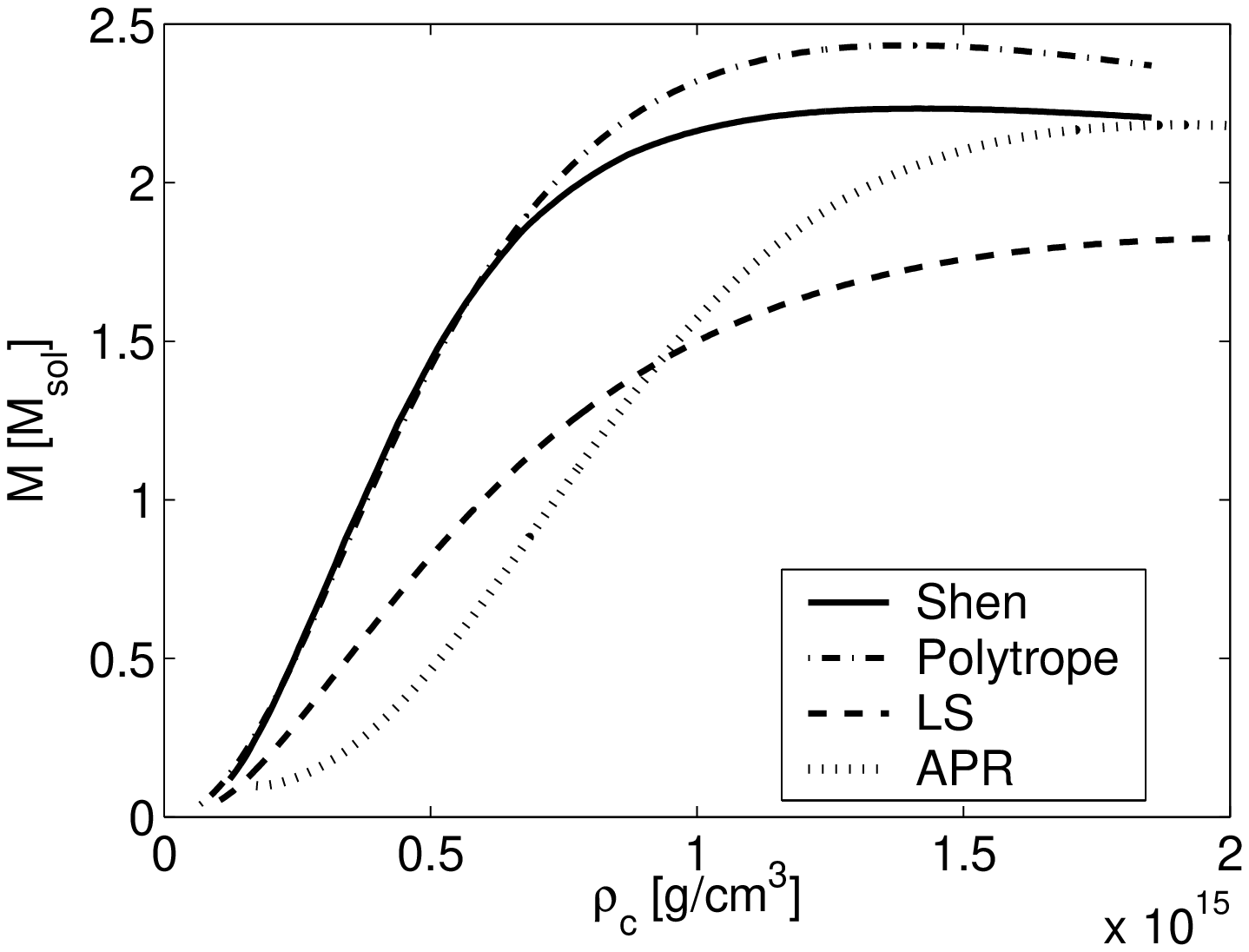}\\
(b)
\includegraphics[width=7.5cm]{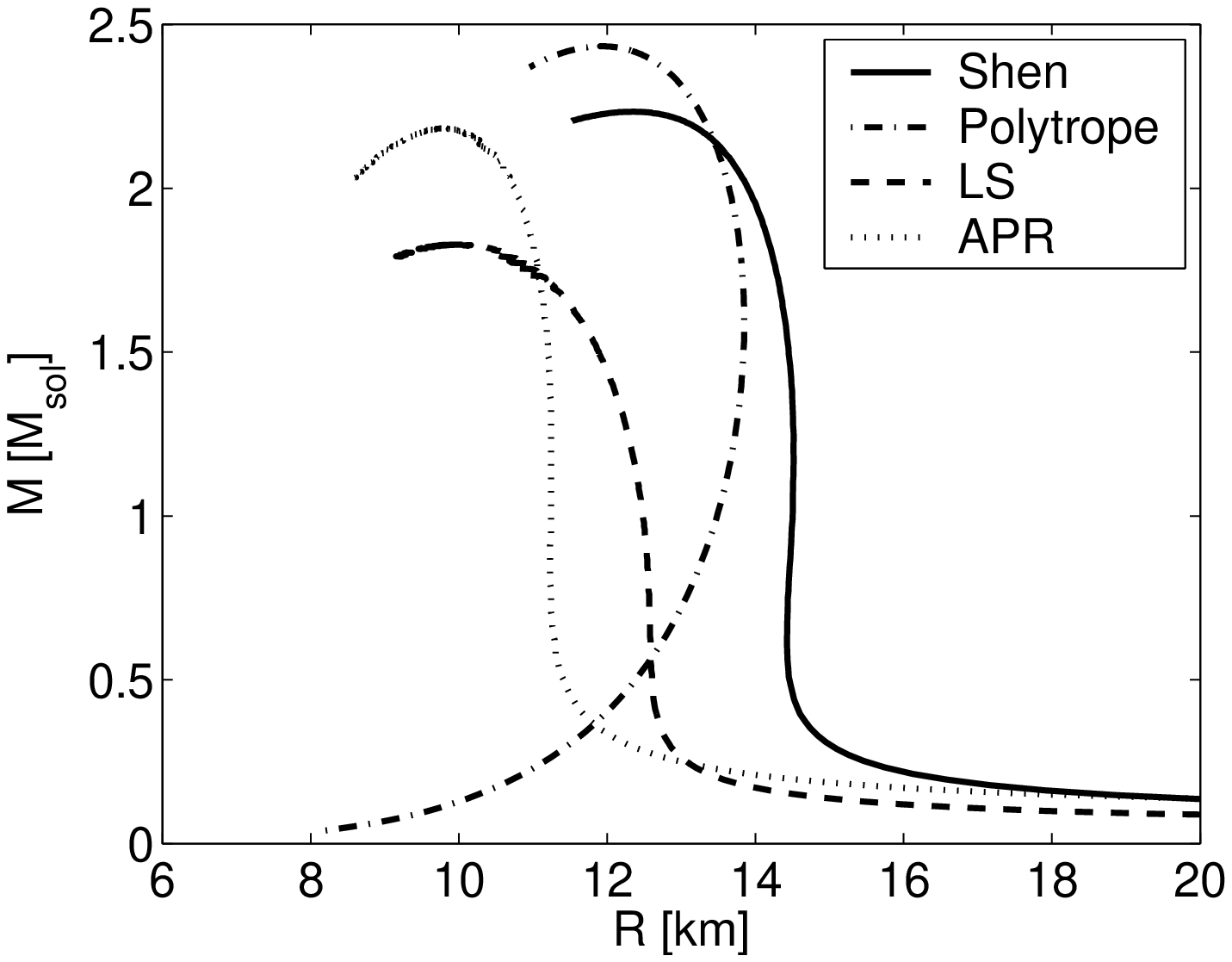}\\
\caption{ADM NS mass $M$ versus central density $\rho_\mathrm{c}$ (panel a) and versus
circumferential radius $R$ (panel b) for the Shen-EoS
(solid) and the LS-EoS (dashed) with $T=0.1$MeV, for the APR-EoS (dotted)
and for the polytropic EoS (dash-dotted). The parameters $K$ and
$\Gamma$ of the latter are chosen such that the $\rho_\mathrm{c}-M$ - relation of the
Shen-EoS is reproduced in a wide range of NS masses up to
$M\sim1.8M_\odot$. The difference between the two EoSs in the $R-M$
relation is due to differences in the low-density regime, which are crucial for determining the NS radius.}
\label{fig:mrprofiles}
\end{center}
\end{figure}

\begin{figure}
\begin{center}
(a)
\includegraphics[width=7.5cm]{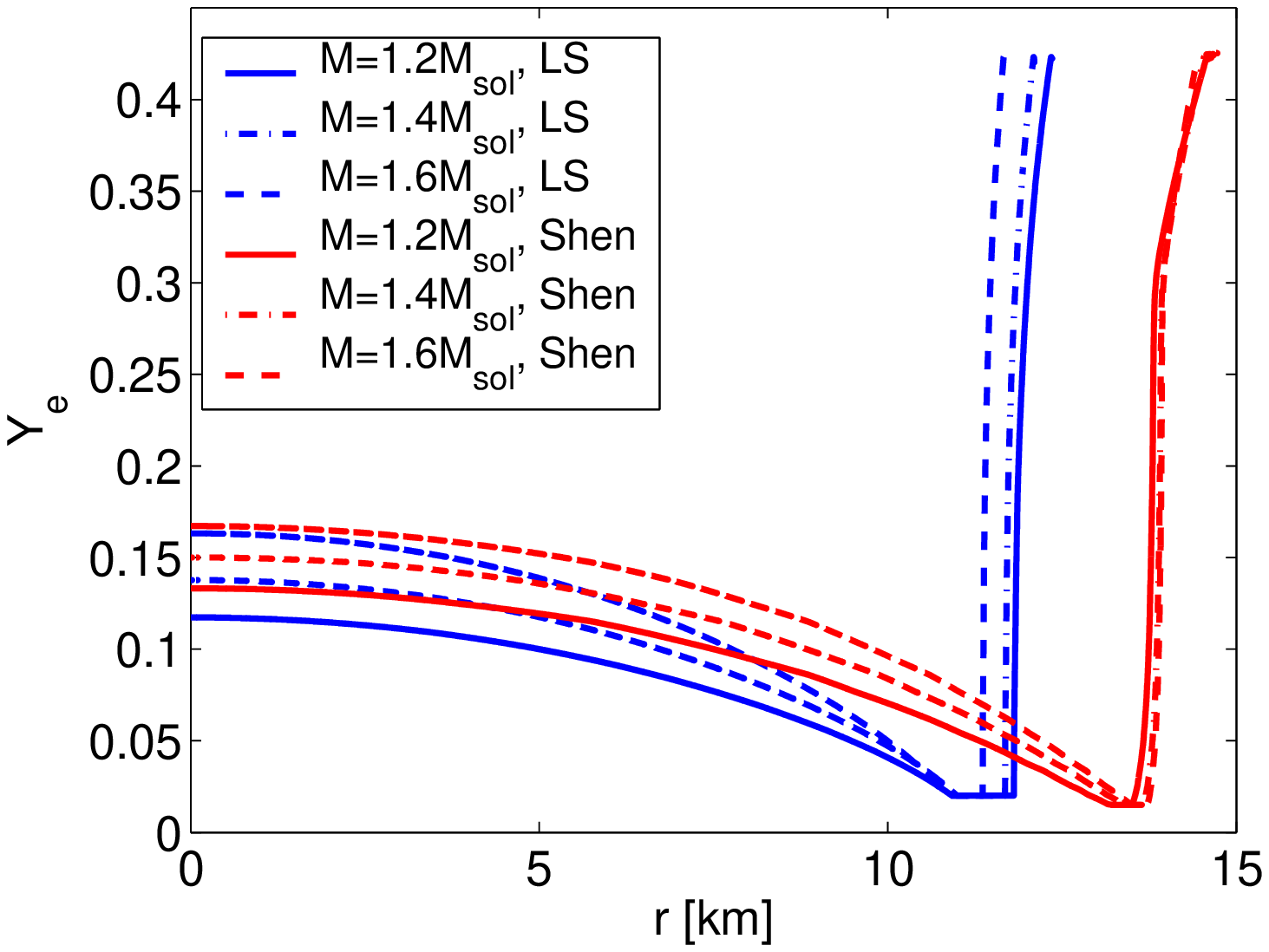}\\
(b)
\includegraphics[width=7.5cm]{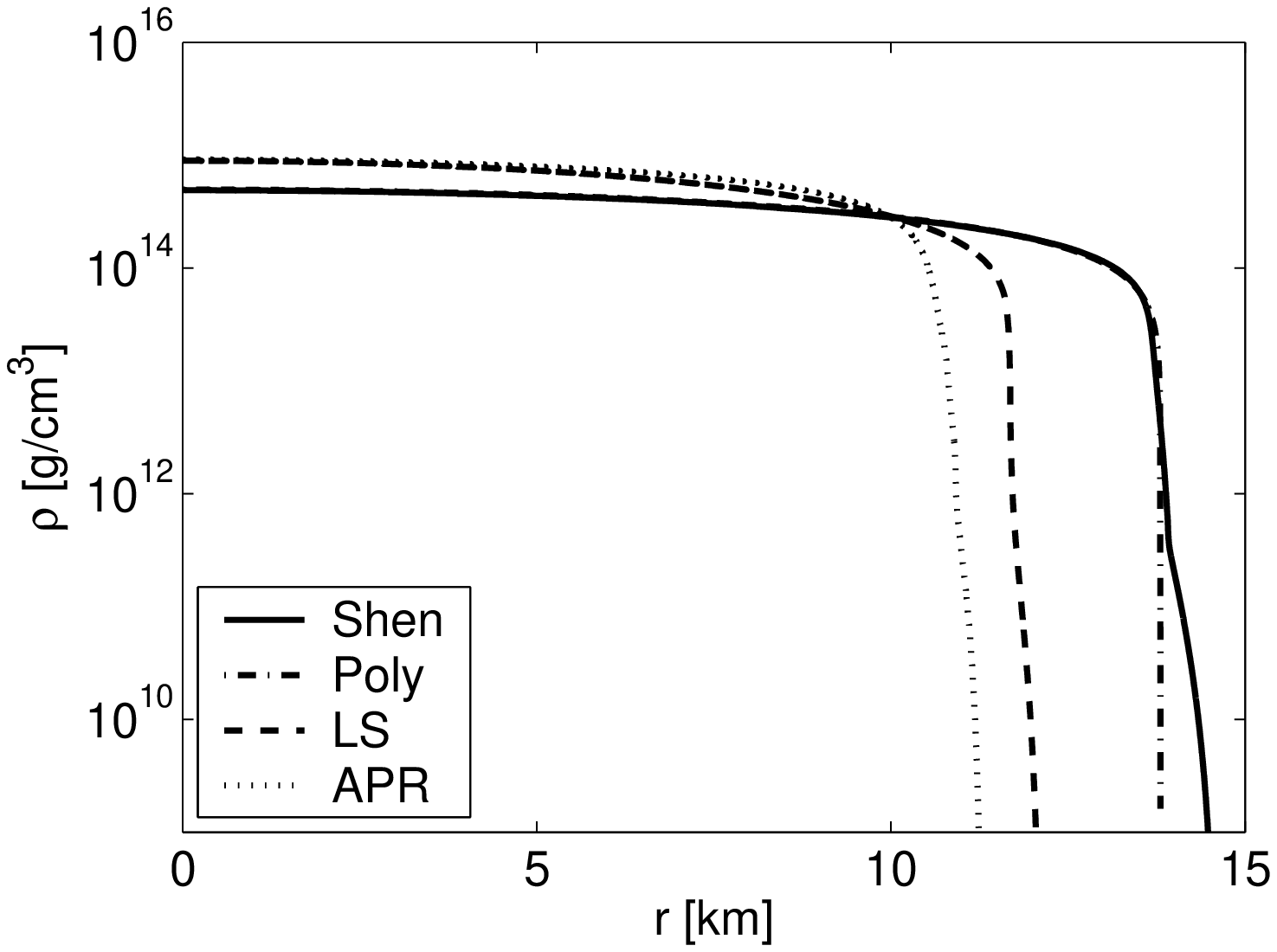}\\

\caption{(a) $Y_{e}$-profiles in NSs with ADM masses of 1.2, 1.4 and 1.6$M_{\odot}$, using both the Shen-EoS and the LS-EoS. The temperature is
set to $T=0.1$MeV and neutrinoless beta-equilibrium is assumed (see
text). Note that a lower limit of $Y_e=0.02$ had to be set in a small
region around nuclear saturation density. (b)
Density profiles for a representative NS with an ADM mass of
$1.4M_\odot$ as computed for the four considered EoSs.}
\label{fig:initNSs}
\end{center}
\end{figure}

The APR-EoS is implemented as in \citet{shibata2006} in order to allow
for a comparison of our results with theirs. The pressure and specific
internal energy are both written as a sum of a thermal and a cold contribution. The
cold component of the specific internal energy $\epsilon_\mathrm{cold}$ is a function
of $\rho$ and is given by an analytic fit formula to the table values
provided by \citet{akmal1998}, who used the variational chain
summation method together with the Argonne $v_{18}$ nucleon-nucleon
interaction model. Three-nucleon-interactions and relativistic boost
corrections are also considered. The cold
pressure component is not directly read from the EoS table but is determined by the thermodynamic relation
$p_\mathrm{cold}=\rho^2d\epsilon_\mathrm{cold}/d\rho$. The thermal pressure component $p_\mathrm{th}$ is given by the ideal-gas law
$p_\mathrm{th}=(\Gamma_\mathrm{th}-1)\rho\epsilon_\mathrm{th}$, where
the specific thermal energy is
given by $\epsilon_\mathrm{th}=\epsilon-\epsilon_\mathrm{cold}$ and
the adiabatic index $\Gamma_\mathrm{th}$
determines the dependence of the pressure on the internal energy at
constant density and is chosen to be $\Gamma_\mathrm{th}=1.5$ or
$\Gamma_\mathrm{th}=2$. Thus, the total pressure $p$ is
determined from the input quantities $\rho$ and $\epsilon$.\\
The APR-EoS has the specific property of
being rather soft around nuclear matter density while being very stiff
at much higher densities (Fig. \ref{fig:p-rho-plot}). This leads to
very compact NSs (Fig. \ref{fig:initNSs}) but still allows for a fairly large maximum NS
masse (Fig. \ref{fig:mrprofiles}). This behaviour is
significantly different from that obtained with the other EoSs. It
will also lead to some peculiarities of the APR models of NS mergers.
 
The ideal-gas EoS is given by the simple relation
$p=(\Gamma-1)\rho\epsilon$, where a polytropic relation
$\epsilon=K/(\Gamma-1)\rho^{\Gamma-1}$ is used to initialize the
internal energy. The parameters
$\Gamma=2.75$ and $K=30000$ (in geometrical units) are chosen such
that the relation between central density and NS mass of the
Shen-EoS is reproduced for a wide range of NS masses (see Fig. \ref{fig:mrprofiles}). This leads to an EoS that
not only closely fits the supernuclear regime of the Shen-EoS at $T\sim
0$ (see Fig. \ref{fig:p-rho-plot}) but also yields very similar NS
density profiles (see Fig. \ref{fig:initNSs}). However, since these EoSs
deviate from each other at subnuclear densities when nuclear forces
become unimportant (see Fig. \ref{fig:p-rho-plot}), we can investigate the importance of this regime
of the EoS during NS coalescence.

We also explore the importance of the non-zero temperature
effects either by varying the parameter $\Gamma_\mathrm{th}$ in the APR-EoS or
by considering cases with the Shen-EoS, where we
artificially neglect the thermal component in the internal
energy. This is done by restricting the full EoS to the
two-dimensional subtable at $T=0$ with density and electron fraction
as remaining independent variables and pressure and internal energy as output quantities (``cold Shen-EoS''). The
energy equation (\ref{eqn:energyeqn2}) is not needed in this case. This reduction has
no influence as long as shocks and viscous heating are absent, i.e. as
long as the fluid evolves adiabatically. In the presence of shock
heating or viscous heating, however, it
corresponds to the extreme case that a very efficient cooling
mechanism extracts immediately the generated entropy and internal
energy. 

\subsection{Estimation of the postmerger torus mass} 

Provided the central, compact merger remnant has a mass above a
certain critical value, it will collapse to a BH, either immediately
after merging or with some delay, after sufficient angular momentum
has been carried away by gravitational radiation or has been
redistributed to matter at larger radii by processes like non-axisymmetric
gravitational interactions, viscosity, and magnetic field
effects \citep{morrison2004}. Therefore, any pressure support from the central remnant
to the surrounding halo gas will vanish and matter will be kept
from infall to the BH mostly by stabilizing rotational support.\\
Consequently, a lower estimate for the mass
remaining in a halo around the BH is given by the condition
\begin{equation}
\label{eqn:criterion}
\hu_{\phi}>j_\mathrm{ISCO},
\end{equation}
i.e., we demand the specific angular momentum $\hu_\phi$ of a fluid
element in the halo (called ``torus'' in the following) to be larger
than the angular momentum at the
innermost stable circular orbit (ISCO) of a Kerr BH with the same mass
and angular momentum as the compact merger remnant.\\
For a Boyer-Lindquist Kerr metric \citep{bardeen1972}, the ISCO
can be analytically determined. In our case, however, we shall use an approximative pseudo-Kerr metric
\citep{grandclement2002} which is both isotropic and conformally flat,
consistent with the coordinates we use for the numerical
simulations. The ISCO can then be found among all circular orbits by
minimizing the specific angular momentum along the radial
coordinate. The detailed calculation is presented in Appendix~\ref{app:disc}.

\section{Initial conditions}
\label{sect:init}

\subsection{Construction of the initial data}

We start our simulations shortly before the tidal instability
occurs and follow the evolution through the merging and torus
formation until either the gravitational collapse of the merger
remnant to a BH sets in or a quasi-stationary
state has formed.\\
At the beginning of the simulations, we have to
provide initial values for the primitive variables $\rho$, $\epsilon$,
and $Y_e$ (in the cases of the Shen-EoS and LS-EoS).

In a first step, we couple the latter two variables uniquely to the
density using the following conditions:
In the Shen-EoS and LS-EoS case, we assume the
two NSs to be cold and in neutrinoless beta-equilibrium, i.e. for the
neutrino chemical potential the relation $\mu_\nu=\mu_n-\mu_p-\mu_e=0$
holds. Together with the condition $T\sim 0$, this fixes uniquely the
electron fraction $Y_e$ and the specific internal energy $\epsilon$
for a given density $\rho$ \footnote{Note that in the case of our tabulated version of the LS-EoS, the
beta-equilibrium condition has no solution in a small region around
nuclear saturation density and at very low temperatures. In this case, we
look for an approximate solution by minimizing $\mu_\nu$ along the
$Y_e$ variable, obtaining $Y_{e}\simeq 0.02$.}. In the APR-EoS, we
implement the ``$T\sim 0$''-condition by requiring
$\epsilon=\epsilon_\mathrm{cold}$. In the case of the ideal-gas EoS,
we use the polytropic relation $\epsilon=K/(\Gamma-1)\rho^{\Gamma-1}$ in order to provide
the initial internal energy. Together with the ideal-gas relation, we
then obtain the familiar expression $p=K\rho^{\Gamma}$ for the initial
pressure. As described in Sect. \ref{sect:eos}, we choose the
parameters $K$ and $\Gamma$ such that the central density-mass
relation of the Shen-EoS for a wide range of NS masses (see
Fig. \ref{fig:mrprofiles}) is reproduced. In all EoS cases we then obtain a unique relation between
density, pressure, and specific internal energy (``1D-EoS'').

In a second step, we use the Oppenheimer-Volkoff-Equation in isotropic
coordinates together with the above described 1D EoSs to obtain radial
NS profiles of the hydrodynamic quantities and the space time metric. We
then generate SPH particle distributions of the two merging NSs in
isolation by placing particles on cylindrical shells around the z-axis
to maintain axisymmetry. The mutual distance between particles is
chosen in order to approximatively obtain a prescribed number of
particles. The particle masses and smoothing lengths are adjusted such
that the particle densities approximate the earlier obtained OV
density profile and such the number of SPH neighbour particles
remains within prescribed limits, typically between 80 and 120 neighbours.

Finally, the two NSs are placed on a circular orbit with a chosen
orbital distance such that the common center of rest mass
$\mathbf{r_{CM}}=\int\rst \mathbf{r}d^3x/\int\rst d^3x$ is located in
the coordinate origin. Here, $\rst$ is the conserved density as defined in
Eq. (\ref{eqn:rhoevol}) and $\mathbf{r}$ is the position vector. The orbital separation
is defined as the coordinate distance between the stellar
centers of rest mass, $\mathbf{r_{CM,i}}=\int_{\mathrm{star~i}}\rst\mathbf{r} d^3x/\int_{\mathrm{star~i}}\rst d^3x$. This system is then
relaxed with the evolution code to an equilibrium configuration by adding a
braking term $\mathbf{f}_\mathrm{brake}$ to the momentum equation,
$$
\mathbf{f}_\mathrm{brake}=C(\hubf-\hubf_\mathrm{eq}),
$$
where $C$ is a chosen small constant, $\hubf$ the current
specific momentum (see Appendix \ref{app:hydro} for definitions) and $\hubf_\mathrm{eq}$ the specific
momentum field to be relaxed to. We set
$$
\hubf_\mathrm{eq}=\hubf_\mathrm{orbit}+\hubf_\mathrm{spin},
$$
where
$$
\hubf_\mathrm{orbit,i}=\mathbf{\hat\Omega}_\mathrm{eq}\times\mathbf{r}_\mathrm{CM,i}
$$
describes the equilibrium orbital motion and
$$
\hubf_\mathrm{spin,i}=\mathbf{\hat\Omega}_\mathrm{i}\times(\mathbf{r}-\mathbf{r}_\mathrm{CM,i})
$$
is the contribution from the stellar rotation. Here, the quantity $\mathbf{\hat\Omega}$ denotes to the angular
velocity related to $\hubf$ (defined in analogy to the angular velocity
$\Omega$ related to the physical velocity $\mathbf{v}$). More
specifically, $\mathbf{\hat\Omega}_\mathrm{i}$ and
$\mathbf{\hat\Omega}_\mathrm{eq}$ refer to the angular velocity related to $\hubf$ of the individual NSs and of the
orbital motion of the binary, respectively. For an irrotating spin setup, we set $\mathbf{\hat\Omega}_\mathrm{i}=0$, for a corotating one
$\mathbf{\hat\Omega}_\mathrm{i}=\mathbf{\hat\Omega}_\mathrm{eq}$, and for a
counterrotating spin setup
$\mathbf{\hat\Omega}_\mathrm{i}=-\mathbf{\hat\Omega}_\mathrm{eq}$. Finally, the
setup with opposite NS rotation means
$\mathbf{\hat\Omega}_1=\mathbf{\hat\Omega}_\mathrm{eq}=-\mathbf{\hat\Omega}_2$.
Now, $\mathbf{\hat\Omega}_\mathrm{eq}$ (and
thus the $\mathbf{\hat\Omega}_\mathrm{i}$) has to be determined such that the mutual gravitational attraction and the
centrifugal forces are in orbital equilibrium. Starting from a guess
value, we do this iteratively during
the relaxation process by increasing $\mathbf{\hat\Omega}_\mathrm{eq}$
if the NSs tend to fall towards each other while decreasing it if the orbital
separation tends to grow. At the same time, the orbital distance
is reset to the initially chosen value.

\begin{figure}
\begin{center}
   \includegraphics[width=7cm]{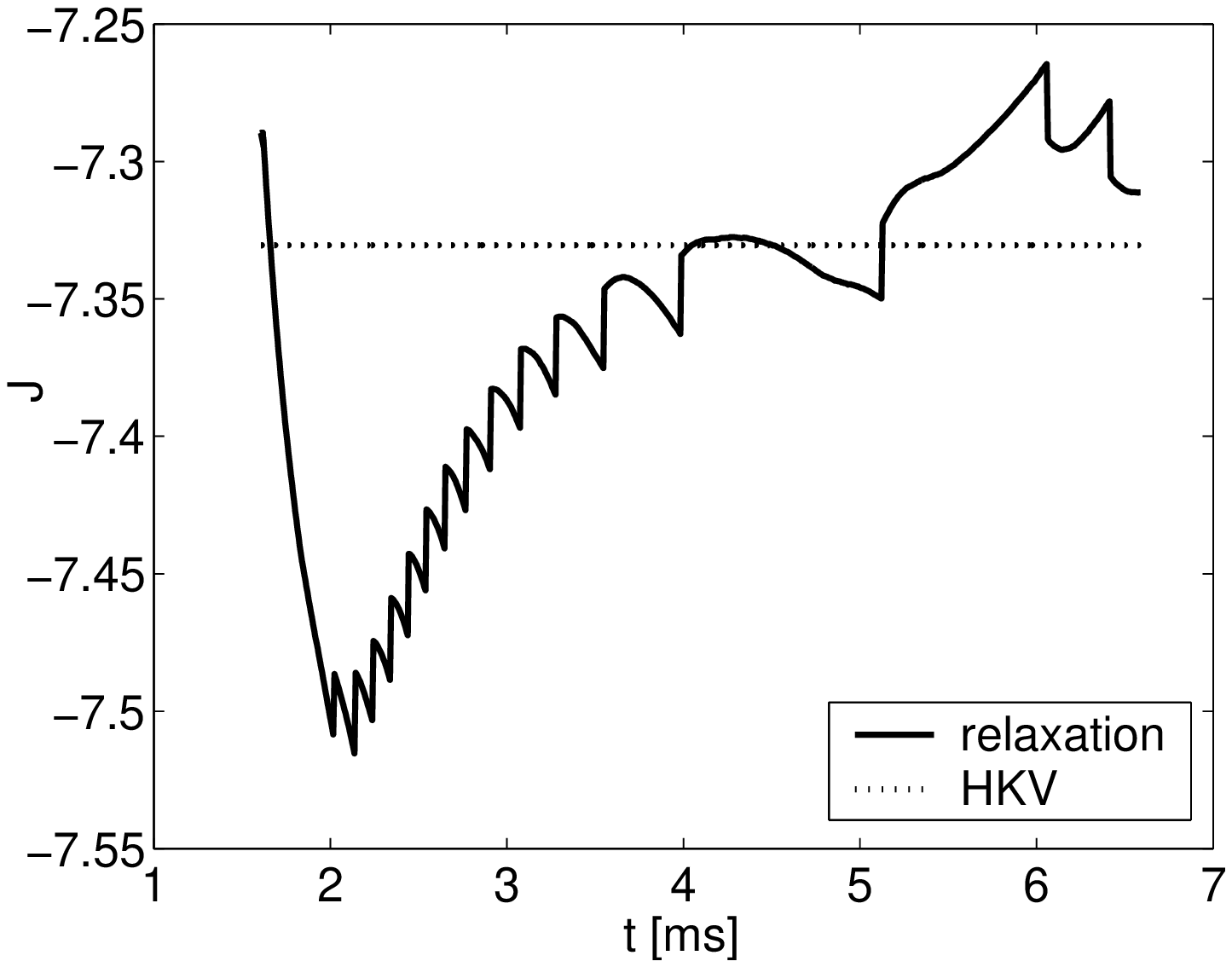}
   \includegraphics[width=7cm]{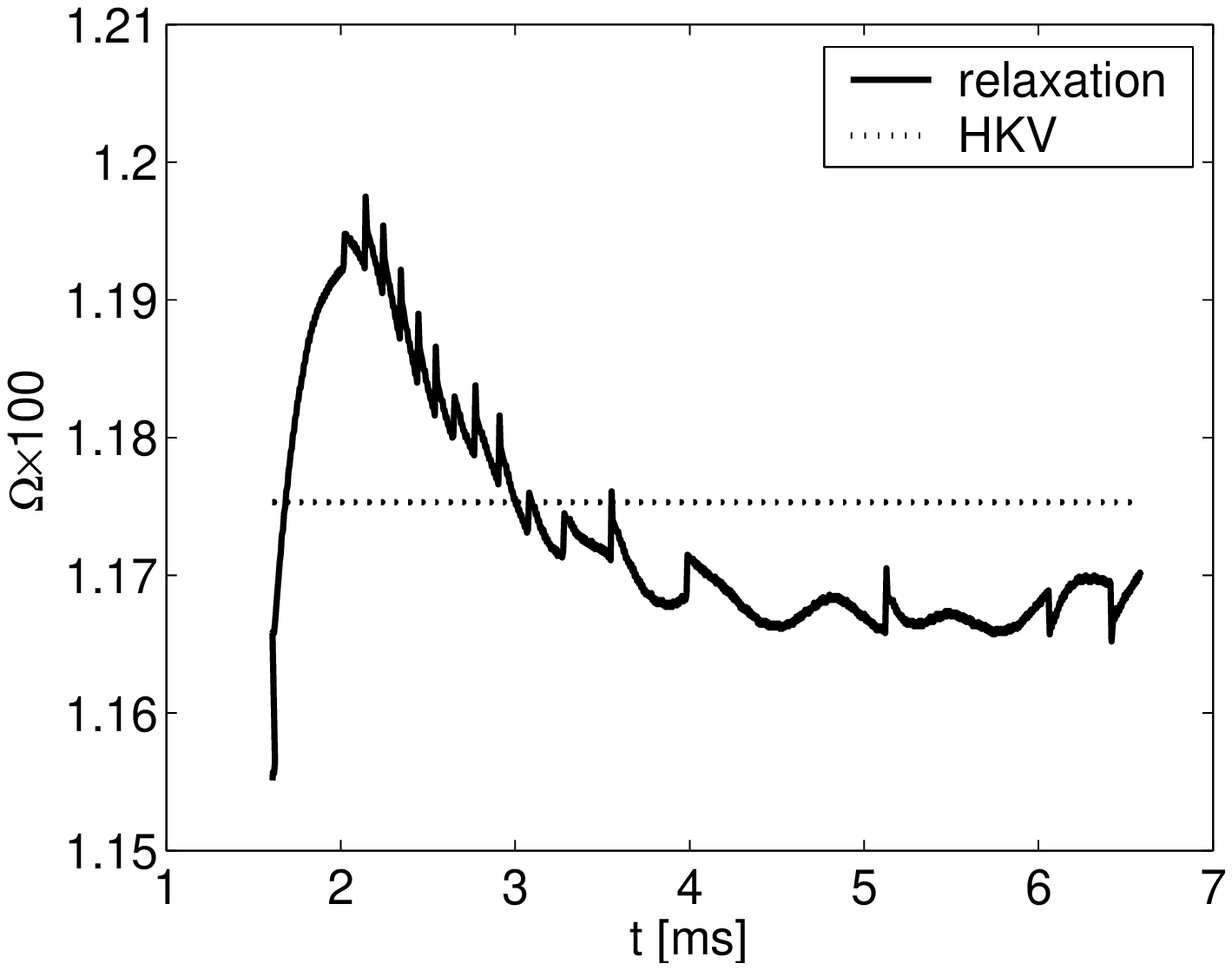}
   \caption{Comparison of the angular momentum (top) and the angular
   velocity (bottom) from our relaxation scheme with results based on
   the helical killing vector (HKV) method. Both quantities are given
   in geometrical units ($G=c=M_\odot=1$).}
   \label{fig:j_relax}
\end{center}
\end{figure}

Using $C=-0.02$, global quantities like the total angular
momentum $J$ and the orbital angular velocity $\Omega$ (both defined in
Appendix \ref{sect:defs}) settle down to stable values within roughly two
orbital periods. Subsequent oscillations at the 1\% level (see
Fig. \ref{fig:j_relax}) are mainly due to readjustments of the orbital
distance during the relaxation process. We usually stop the
calculation after about four orbital
periods and read off the initial data at the end.\\ 
The definition of an irrotational
velocity field by $\mathbf{\hat\Omega}_\mathrm{i}=0$ is motivated from
Newtonian physics. It satifies the equations used by the helical
killing vector (HKV) method \citep{shibatairrotational1998,gourgoulhon2001} to Newtonian order. To
investigate the error we introduce with our approximative definition
of irrotationality, we compare the obtained results to
quasi-equilibrium models based on the HKV method
\citep{uryu2000,gourgoulhon2001}. We typically find devitations in the
angular momentum $J$ and the angular velocity $\Omega$ of the order of
1\% (Fig. \ref{fig:j_relax}). Note that we could also have used
the physical velocity $\mathbf{v}$ rather than the specific momentum
$\hubf$ for the definition of the equilibrium velocity field since
the latter two quantities are equal to Newtonian order. However, the
use of the specific momentum yields a much better agreement in $J$ and
$\Omega$ with results obtained with the HKV method.

\subsection{Initial models}

In order to study the influence of the various parameters that
characterize the initial model and to determine the subsequent merger dynamics, we vary
the individual NS masses $M_1$ and $M_2$, the NS spin setup, and the EoS. In Table \ref{tab:inittable} we summarize the
describing parameters of our models.

We consider both equal-mass models, i.e. with a
mass ratio $q=M_1/M_2=1$, and unequal-mass models in order to
investigate the consequences of $q<1$. A mass ratio near unity seems to be
favoured by observations, but population synthesis models suggest that
beside a population component of systems with nearly equal masses also a significant fraction of binaries with
$q$ as low as ~0.55 might exist \citep{bulik2003}.

Many of our models are chosen with zero NS spin (``irrotating''), which is
generally considered to be a good approximation to reality because the viscous
timescale is too large to allow for tidal locking during inspiral and
because the orbital period near the ISCO of
$~2$ms is considerably smaller than typical NS spin periods
\citep{kochanek1992, bildsten1992}. Observed NS spin periods in double
NS binaries are indeed roughly ten times larger \citep{stairs2004}.
In view of the large influence of the NS spins on the merger, we still
think that a set of representative models with non-zero NS spins is
worth being considered. 

\begin{table*}
\begin{center}
\caption{Characteristic parameters and data of our considered models. $M_1$ and $M_2$
denote the individual gravitational masses in isolation while $M_\mathrm{sum}=M_1+M_2$
stands for the sum of the two (Note that the total gravitational mass $M$ is slightly smaller than $M_\mathrm{sum}$ because $M$
also involves the negative gravitational binding energy.). $M_0$ is the total baryonic mass, whereas $q=M_1/M_2$ and
$q_{M}=M_{0,1}/M_{0,2}$ are the gravitational and baryonic mass ratio, respectively. `Shen' stands for the full, finite 
temperature Shen-EoS, `Shen\_c' for the restriction to zero
temperatures, `LS' denotes the LS-EoS, `ideal' means
ideal-gas EoS, and, finally, `APR15' and `APR2' refer to the APR-EoS
extended using $\Gamma_\mathrm{th}=1.5$ and $\Gamma_\mathrm{th}=2$, respectively. The
NS spin states are denoted as `irrot' for irrotating,
`corot' for corotating, `c\_rot' for counterrotating, `oppo' for
oppositely oriented spins and `tilted' for a spin orientation inclined
relative to the orbital spin. More specifically, we choose in model
S1414t1 the angular velocity vectors $\mathbf{\hat\Omega}_1=0$ and
$\mathbf{\hat\Omega}_2=0.041*(0,1,0)$ (in geometrical units), which
corresponds to a spin period of the second NS of about  1 ms, while in
model S1414t2, we set $\mathbf{\hat\Omega}_1=0.01*(0,1,1)$ and
$\mathbf{\hat\Omega}_2=0.01*(0,-1,1)$.
$T_\mathrm{max}$ is the maximum temperature reached in the system
during the whole evolution. It is obtained by averaging the particle
temperatures on a grid with cells of $1.5$km side
length. The rotation parameters $a_\mathrm{ini}=J_\mathrm{ini}/M_\mathrm{ini}^2$,
 $a_\mathrm{system}$, and $a_\mathrm{remnant}$ refer to the initial
binary model, the entire system right after merging, and to only the
compact merger remnant at that time. Finally, $M_\mathrm{torus}$ and $M_\mathrm{ej}$ denote
the (baryonic) torus mass and ejecta mass about 5ms after merging when
a quasi-stationary state has formed.}

\label{tab:inittable}
\begin{tabular}{c|c|c|c|c|c|c|c|c c|c|c|c|c|c|c}
Model & $M_1$ & $M_2$ & $M_\mathrm{sum}$ & $M_0$ & $q$ & $q_{M}$ &EoS& Spins&
&$T_\mathrm{max}$&$a_\mathrm{ini}$&$a_\mathrm{system}$&$a_\mathrm{remnant}$&$M_\mathrm{torus}$&$M_\mathrm{ej}$\\
\hline
Units & $M_\odot$ & $M_\odot$ & $M_\odot$ & $M_\odot$ & & $ $ & & & &
MeV & & & & $M_\odot$&$\hspace{-0.1cm}10^{-3}M_\odot\hspace{-0.1cm}$\\ 

\hline
S1414 & 1.4 & 1.4 & 2.8& 3.032 & 1.0 & 1.0 &Shen&irrot&$\times
\times$&52&0.96&0.91&0.88&0.04&$1.5$ \\
S138142 & 1.38 & 1.42 &2.8& 3.032 & 0.97 & 0.97&Shen&irrot&$\times
\times$&50&0.96&0.90&0.87&0.04&$3$\\
S135145 & 1.35 & 1.45 &2.8& 3.034 & 0.93 & 0.93&Shen&irrot&$\times
\times$&50&0.96&0.90&0.86&0.07&$5$\\
S1315 & 1.3 & 1.5 &2.8& 3.037 & 0.87 &0.86&Shen&irrot&$\times
\times$&50&0.96&0.90&0.84&0.14&$6$\\
S1216 & 1.2 & 1.6 &2.8& 3.039 & 0.75 &0.73&Shen&irrot&$\times
\times$&54&0.95&0.89&0.76&0.21&$12$\\
S1515 & 1.5 & 1.5 &3.0& 3.274 & 1.0 &1.0&Shen&irrot&$\times \times$&67&0.94&0.89&0.86&0.04&-\footnotemark[3] \\
S1416 & 1.4 & 1.6 &3.0& 3.274 & 0.88 &0.86&Shen&irrot&$\times
\times$&55&0.94&0.89&0.84&0.14&7 \\
S1317 & 1.3 & 1.7 &3.0& 3.279  & 0.76 &0.75&Shen&irrot&$\times
\times$&65&0.94&0.88&0.78&0.20&8 \\
S119181& 1.19 & 1.81 & 3.0 & 3.289 & 0.66 & 0.63 &Shen&irrot&$\times \times$&57&0.91&0.87&0.72&0.24&25\\
S107193& 1.07 & 1.93 & 3.0 & 3.306 & 0.55 & 0.52 &Shen&irrot&$\times \times$&70&0.87&0.84&0.69&0.26&45 \\
S1313 & 1.3 & 1.3 &2.6& 2.800 & 1.0 &1.0&Shen&irrot&$\times \times$&52&0.99&0.93&0.89&0.06&4\\
S1214 & 1.2 & 1.4 &2.6& 2.799 & 0.86 &0.85&Shen&irrot&$\times \times$&43&0.98&0.93&0.82&0.19&-\footnotemark[3]\\
S1115 & 1.1 & 1.5 &2.6& 2.807 & 0.73 &0.71&Shen&irrot&$\times \times$&50&0.99&0.92&0.76&0.22&13\\
\hline
C1216 & 1.2 & 1.6 &2.8& 3.039 & 0.75 &0.73&Shen\_c&irrot&$\times \times$&-\footnotemark[1]&0.95&0.90&0.75&0.28&5\\
C1315 & 1.3 & 1.5 &2.8& 3.037 & 0.87 &0.86&Shen\_c&irrot&$\times \times$&-\footnotemark[1]&0.96&0.91&0.81&0.20&3\\
P1315 & 1.3 & 1.5 & 2.8& 3.064 & 0.87 & 0.86 &ideal&
irrot&$\times \times$&-\footnotemark[1]&0.96&0.90&0.84&0.14&8\\
\hline
LS1414 & 1.4 & 1.4 & 2.8& 3.077 & 1.0 & 1.0 &LS&irrot&$\times
\times$&175&0.96&0.86&0.85\footnotemark[3]&0.02\footnotemark[3]&2\\
LS1216 & 1.2 & 1.6 &2.8& 3.087 & 0.75 &0.73&LS&irrot&$\times
\times$&-\footnotemark[2]&0.95&0.85\footnotemark[2]&0.78\footnotemark[2]&0.19\footnotemark[2]&-\footnotemark[2]\\
A151414 & 1.4 & 1.4 & 2.8& 3.032 & 1.0 & 1.0 &APR15&irrot&$\times
\times$&-\footnotemark[1]&0.97&0.84&0.81\footnotemark[3]&0.06\footnotemark[3]&2\\ 
A21414 & 1.4 & 1.4 & 2.8& 3.032 & 1.0 & 1.0 &APR2&irrot&$\times
\times$&-\footnotemark[1]&0.97&0.84&0.83\footnotemark[3]&0.04\footnotemark[3]&2\\
\hline
S1414co & 1.4 & 1.4 & 2.8& 3.032 & 1.0 & 1.0 &Shen&corot&$\uparrow
\uparrow$&40&1.05&0.98&0.86&0.21&1 \\
S1414ct & 1.4 & 1.4 & 2.8& 3.032 & 1.0 & 1.0 &Shen&c\_rot&$\downarrow
\downarrow$&74&0.88&0.85&0.83&0.02&2.5\\
S1414o & 1.4 & 1.4 & 2.8& 3.032 & 1.0 & 1.0 &Shen&oppo&$\uparrow
\downarrow$&46&0.97&0.92&0.90&0.07&2 \\
S1414t1 & 1.4 & 1.4 & 2.8& 3.032 & 1.0 & 1.0 &Shen&tilted&$\leftarrow
\times$&59&0.99&0.95&0.85&0.20&28\\
S1414t2 & 1.4 & 1.4 & 2.8& 3.032 & 1.0 & 1.0 &Shen&tilted&$\nearrow
\nwarrow$&39&1.05&0.98&0.86&0.23&-\footnotemark[3]\\
S1216co & 1.2 & 1.6 &2.8& 3.039 & 0.75 &0.73&Shen&corot&$\uparrow
\uparrow$&48&1.03&0.98&0.78&0.28&33\\
S1216ct & 1.2 & 1.6 &2.8& 3.039 & 0.75 &0.73&Shen&c\_rot&$\downarrow
\downarrow$&73&0.87&0.84&0.74&0.18&14\\
\hline
\end{tabular}
\end{center}
\end{table*}

\section{Results}
\label{sect:results}

In the further analysis and discussion of our models, with have
normalized the time axis such that $t=0$ coincides with the time when
the gravitational-wave luminosity during merging becomes maximal. This
allows for a better comparison of the merger and postmerger evolution
of different models.

\subsection{Merger dynamics}

\footnotetext[1]{EoS does not contain temperature information.}
\footnotetext[2]{Quantity still changes when the collapse to
a BH sets in immediately after merging.}
\footnotetext[3]{Quantity still increases at the end of the run.} 

We start our simulations somewhat outside the ISCO, i.e. at an
orbital separation where the binary is still stable. In our
equilibrium initial models we neglect the
radial component of the orbital velocity. \citet{miller2004} suggests
that this leads to a systematic error in the inspiral dynamics and the
time until merging. We have performed test calculations with different initial
orbital distances and with artificially reduced initial
angular momentum and backreaction strength. They indeed show a different
inspiral behaviour, but the merger and postmerger phases depend very
weakly on the inspiral dynamics. \citet{shibata2002} arrived at a
similar conclusion. Since we concentrate in this work on the merger and
postmerger dynamics and on the final outcome, we can safely rely on
this approximation. In the following subsections, we are going to investigate the impact
of the three parameters -- the NS mass ratio, the nuclear EoS, and the
initial NS spins -- which largely determine the merger dynamics, the
remnant properties, and the possible formation of a torus.

\subsubsection{Different mass ratios}

\begin{figure*}
\begin{center}
  \begin{minipage}[t]{0.47\linewidth}
   \includegraphics[width=7.5cm]{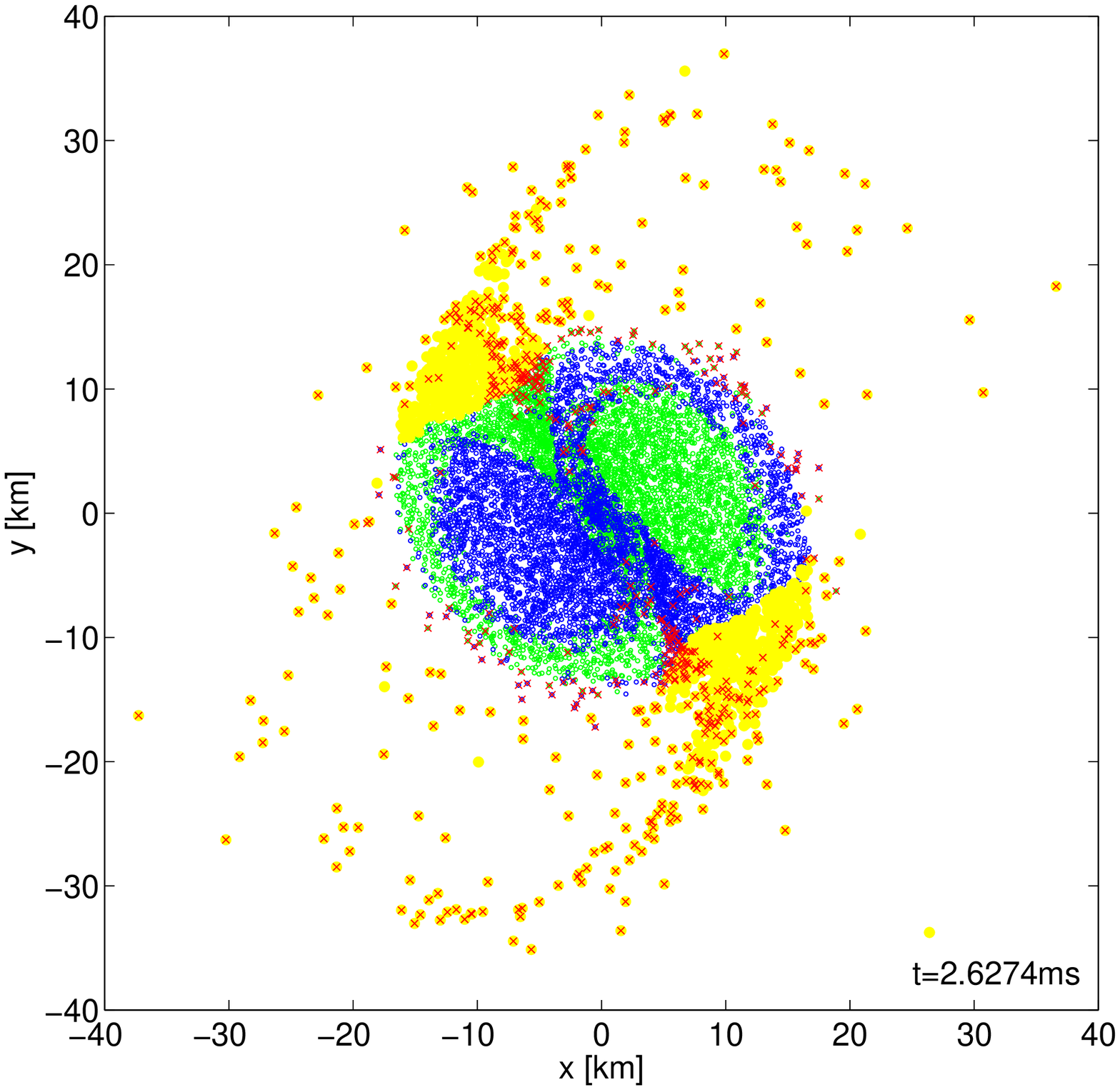}
	(a)
\end{minipage}
  \begin{minipage}[t]{0.47\linewidth}
   \includegraphics[width=7.5cm]{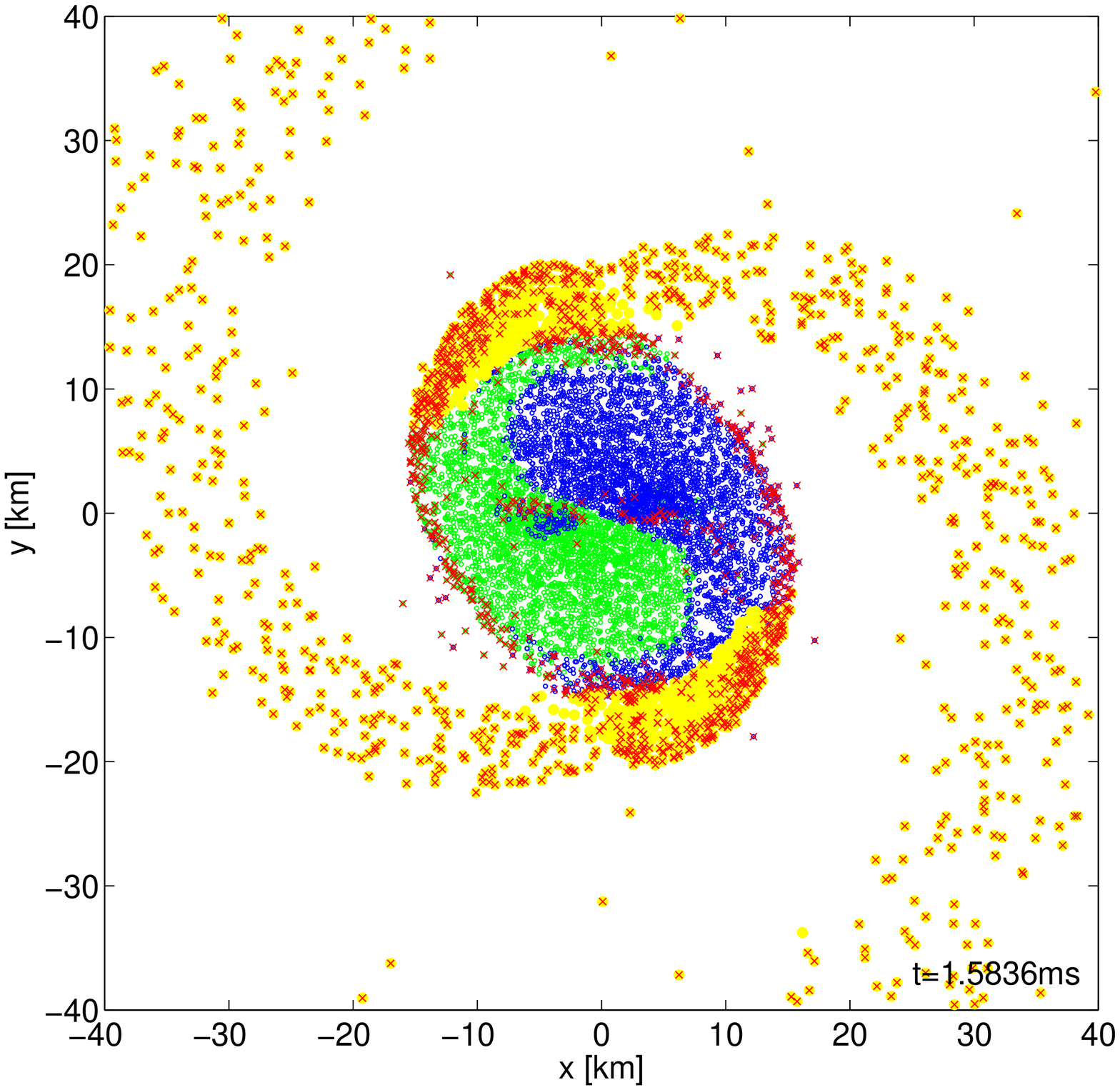}
	(b)
\end{minipage}
\\
  \begin{minipage}[t]{0.47\linewidth}
   \includegraphics[width=7.5cm]{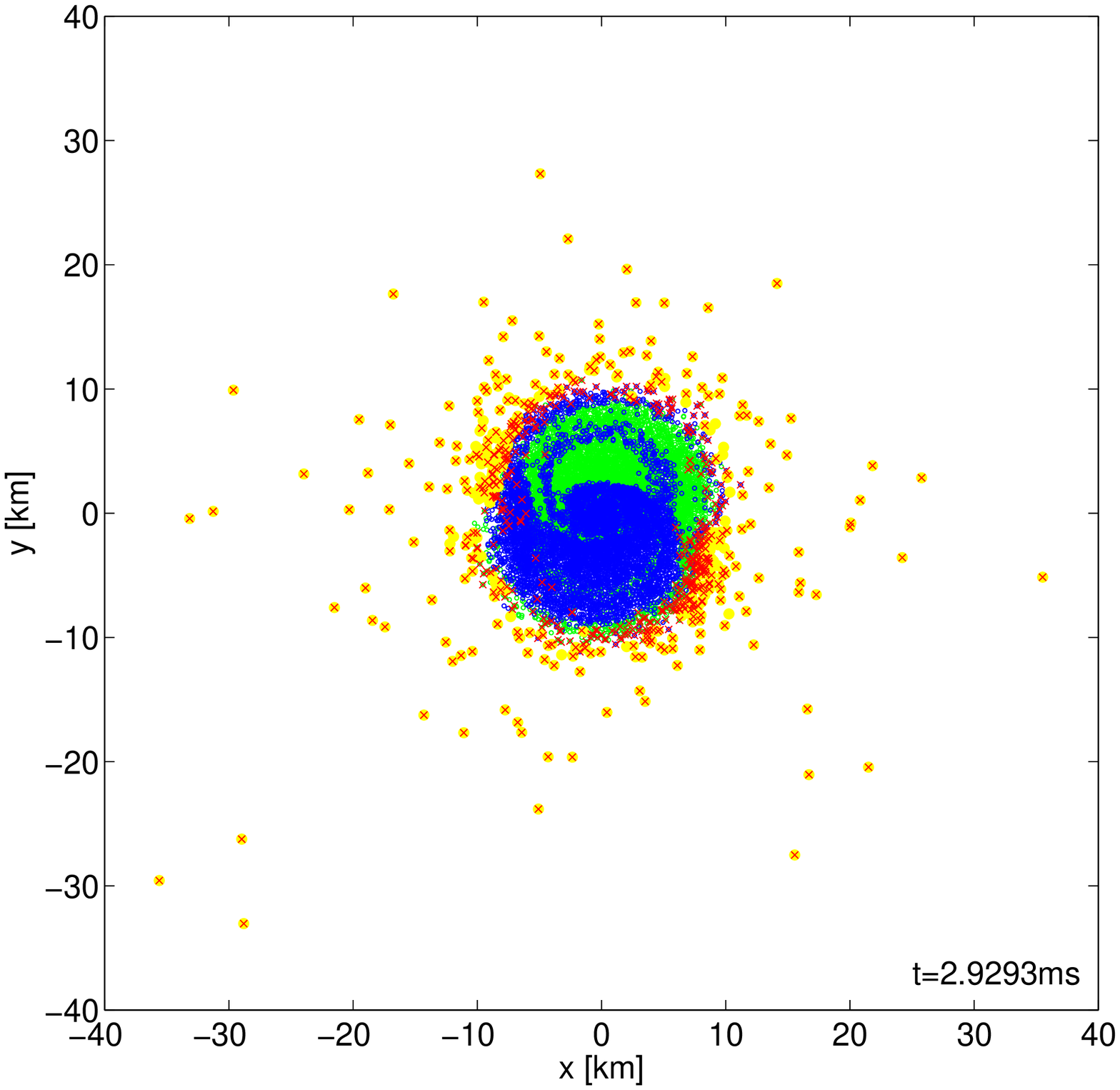}
	(c)
\end{minipage}
  \begin{minipage}[t]{0.47\linewidth}
   \includegraphics[width=7.5cm]{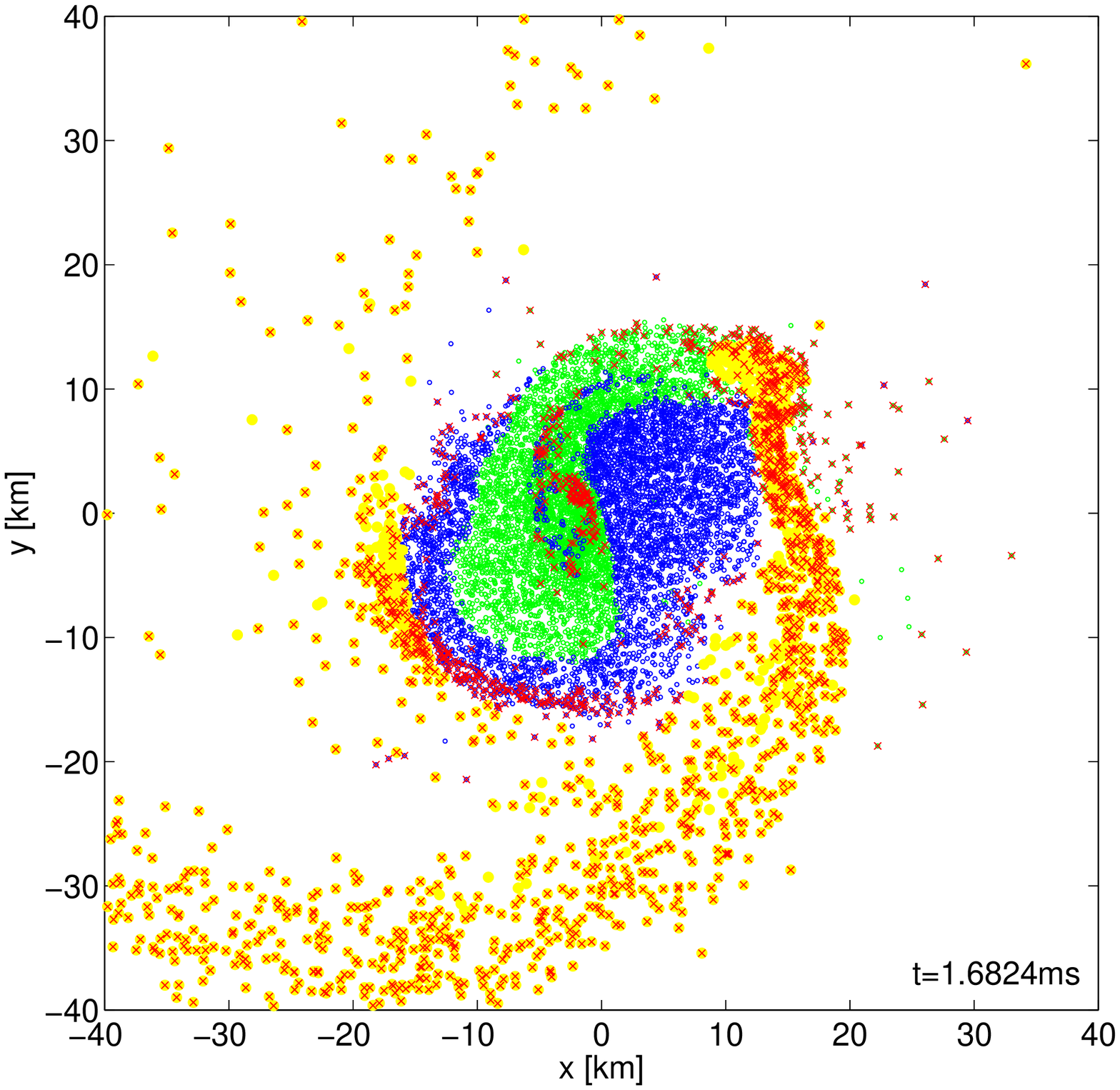}
	(d)
\end{minipage}
\caption{Characteristic snapshots of four representative
models. Plotted is every 10th SPH particle around the equatorial
plane. The matter of the two NSs is color-coded by green and blue
particles. Yellow circles represent particles which currently fulfill
the torus criterion of Eq.(\ref{eqn:criterion}) whereas particles
ultimately ending up in the high-angular-momentum torus are plotted in red. By definition, the yellow and red particles
coincide at the end of the run. Panel (a) shows the development of
secondary spiral arms in model S1414. Panel (b) shows the large,
primary spiral arms in model S1414co. The two clumps of red particles
at the remnant surface are going to form secondary spiral arms which
transport these particles into the torus. Panel
(c) shows a postmerger snapshot of model LS1414. The merger remnant
in much more compact and no secondary spiral arms are formed. As a
consequence, the torus is much smaller and develops only on a longer
timescale. In panel (d), we see the formation of the large primary
spiral arm in model S1216. Similar to panel (b), a secondary spiral
arm begins to form in the area at the remnant surface where the red particles cluster.}
\label{fig:particles}
\end{center}
\end{figure*}

\begin{figure*}
\begin{center}
   \includegraphics[width=18cm]{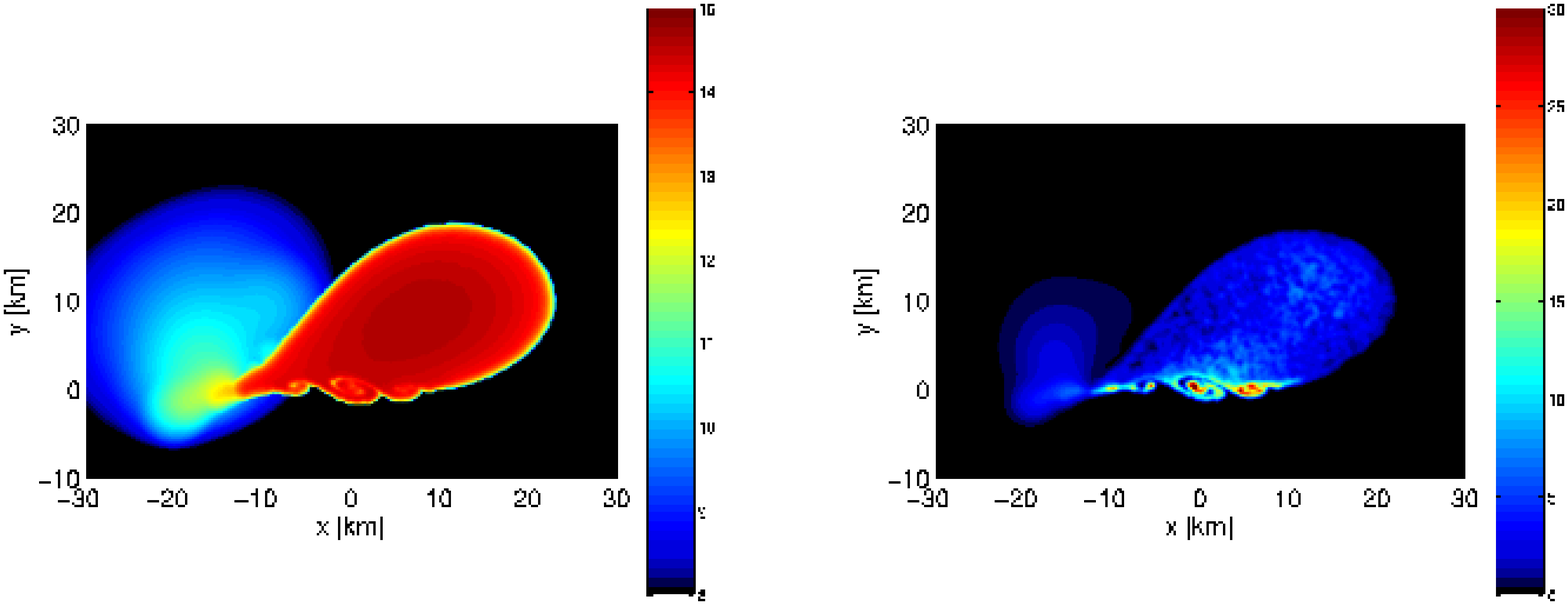}
   \caption{Visualisation of the Kelvin-Helmholtz-unstable shear
   interface and of the associated shock-heating and viscous shear heating
   during merging of the two NSs in model S1414 at $t\simeq
   0$. Plotted is the density (left; logarithmic scale, in g/cm$^3$) and temperature
   (right; linear scale, in MeV) distribution of the matter originating from one of
   the NSs of this model.}
\label{fig:KH}
\end{center}
\end{figure*}

In models with a mass ratio considerably different
from unity, e.g. in models S107193, S1216 or S1317, the less massive star is tidally stretched and disrupted, forming a long spiral
arm which is accreted to a large part on its more massive
companion. This leads to a non-axisymmetric, differentially rotating
central object (called ``remnant'' in the following). The core of the
central remnant forms from the more massive NS while the accreted material from
the less massive partner is wound up to end in the outer shells of the
remnant. Due to a `sling-shot-effect' in the spiral arm, a
significant amount of angular momentum is carried from the remnant to
larger radii such that the material sitting in the spiral arm tip
manages to escape accretion and to contribute to the main part of a
future torus \citep[for a detailed discussion, see][]{oechslin2006a}. A
small fraction of this material can even become gravitationally
unbound and might escape from the system (see
Sect. \ref{sect:ejecta}). This angular momentum transport -- and also
the fact that asymmetric systems already have a smaller angular
momentum in the pre-merging quasiequilibrium
stage -- leads to smaller spin parameters $a_\mathrm{remnant}$ of the
remnant in asymmetric systems (see Table \ref{tab:inittable}). In
some systems, this angular momentum difference leads to an immediate
collapse to a BH after the merger (model LS1216) whereas the symmetric
model (LS1414) with the same EoS and spin state collapses not until
about 10ms after merging. During the postmerger evolution, strong radial and non-axisymmetric
oscillations of the central remnant periodically lead to the formation of smaller, secondary
spiral arms and to successive ejection of material into the torus (see Fig. \ref{fig:particles}).

In contrast, in case of a mass ratio near unity as, e.g., in models S1414
or S138142, the two partners smoothly plunge into each other without
any disruption and without the formation of a primary spiral arm. The resulting
central object has a twin-core structure formed from the two
original cores (Fig. \ref{fig:particles}, panel a). They are kept
mostly intact during merging and
lead to a rotating bar-like structure in the remnant center, which
drives the formation of secondary spiral arms after the dynamical
merger phase and, which leads to ejection of material into the torus.
The bar-shaped core slowly transforms during the later evolution into
a rotationally flattened high-density core. The mixing of material between the
two initial NSs in the remnant is mainly
taking place at the turbulent contact interface (see Fig. \ref{fig:KH}). Since the primary spiral arm is
absent, a much smaller torus results after merging. Note, however, that primary
spiral arms and large torus masses may still emerge if the initial NS
spin setup is favourable, e.g. in case of corotating binaries as in
model S1414co.

\subsubsection{Different neutron star spins}
\label{sect:results-spins}

The NS spins determine the velocity jump across the contact
interface at merging. This
velocity discontinuity leads to a Kelvin-Helmholtz (KH)-unstable vortex
sheet between the two merging NSs and to shock heating (and heating by
numerical viscosity) along the
contact interface. In models with a large velocity jump as in
irrotating or even counterrotating models, we indeed see large
KH-vortices and strong heating (Fig. \ref{fig:KH}). In contrast, in corotating models like
S1414co, there is only a small velocity jump which
does not result from the spinning NSs but from the Coriolis
force deflecting the infalling gas in the tidal tips into azimuthal
direction shortly before merging. Consequently,
KH-vortices still develop, but on a longer timescale and with reduced
shock heating and viscous heating.

\begin{figure}[h]
\begin{center}
   \includegraphics[width=7cm]{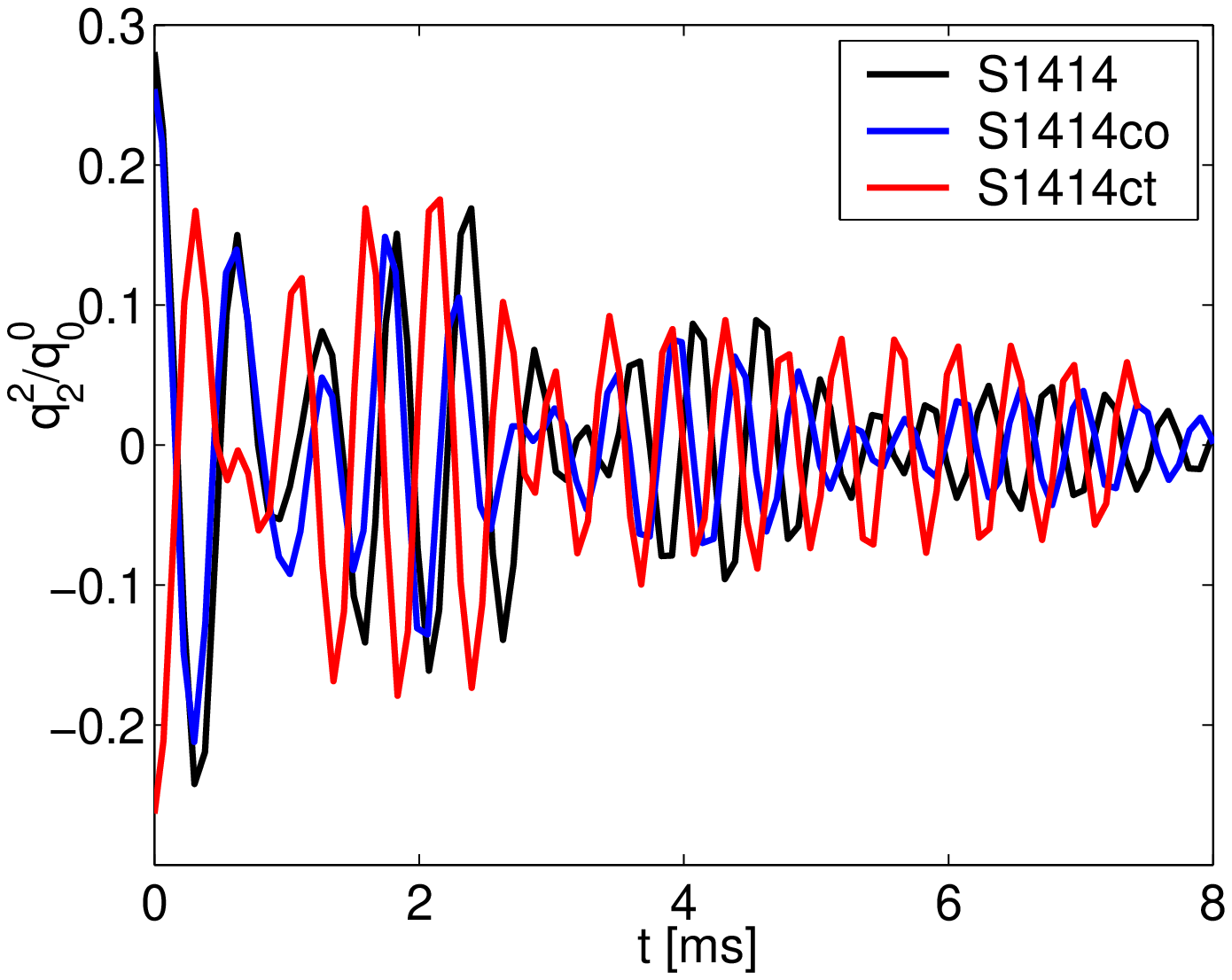}
(a)	
   \includegraphics[width=7cm]{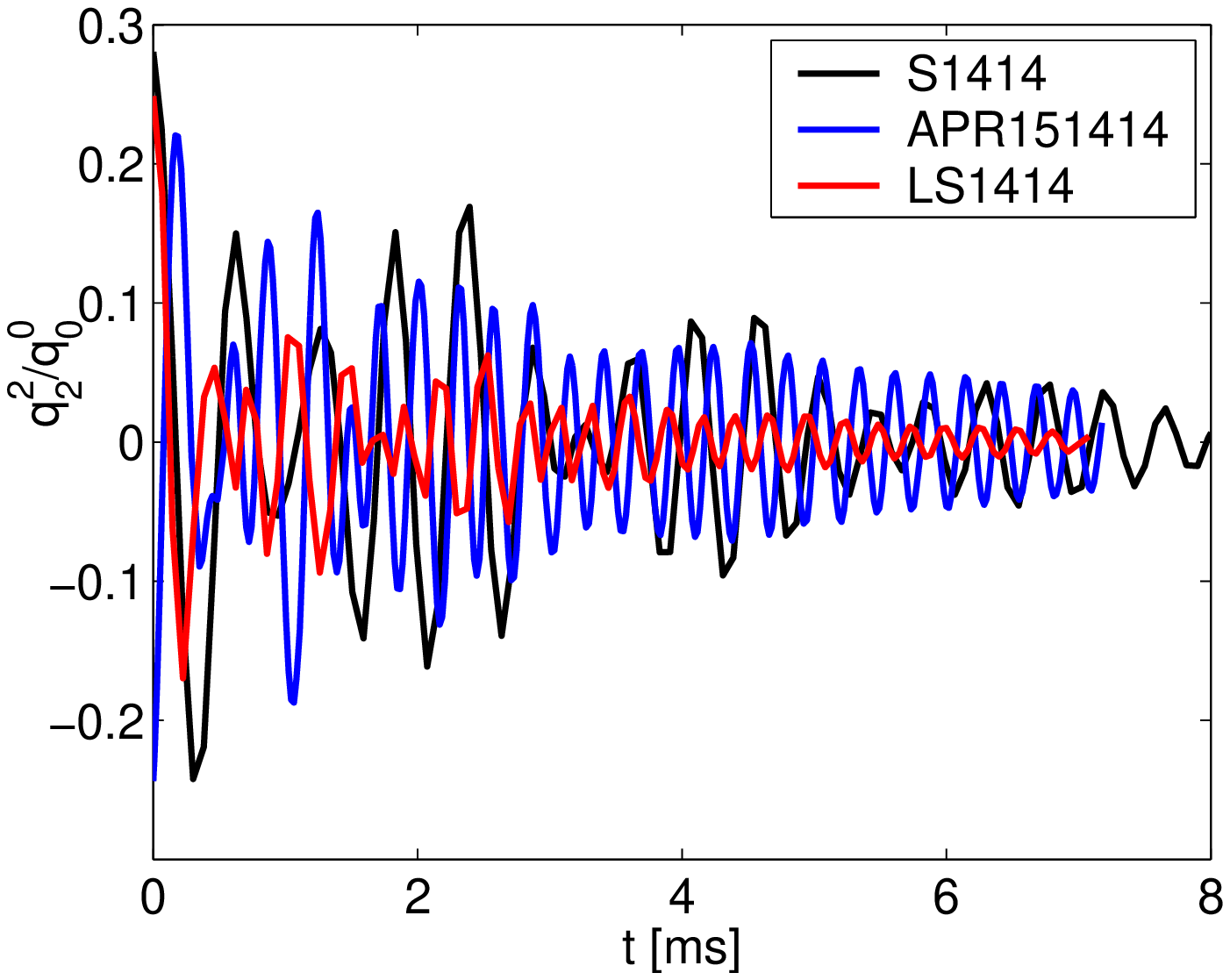}
(b)
   \caption{(a) Plotted is the normalized $l=m=2$-mode amplitude of
the remnant rest mass
   distribution as defined in Sect. \ref{sect:results-spins} for models S1414, S1414co and S1414ct. The evolution
of the amplitude,
   which is a measure for a deviation from axisymmetry, is significantly
   larger at later times in the counterrotating model than in the other two
   cases. (b) The same mode amplitude, but
   for the $1.4-1.4M_\odot$-models with the Shen-, LS- and APR-EoSs. The strength of non-axisymmetry increases
with the stiffness of the supernuclear EoS.  Note that the normalized $l=m=2$-mode
is not sensitive to the compactness of the remnant.}
   \label{fig:q22mode}
\end{center}
\end{figure}

A larger velocity jump and thus a stronger shear flow at the interface
is also associated with stronger non-axisymmetric deformation of the central remnant. This can be seen, e.g., by
considering the normalized $l=m=2-$mode amplitude $q^2_2/q^0_0=\int\rst
(x^2-y^2)d^3x/\int\rst (x^2+y^2+z^2)d^3x$ which is an
oscillating function whose amplitude can be considered as a measure
for the deviation from non-axisymmetry. In Fig. \ref{fig:q22mode}, we see that the non-axisymmetry, as measured by the above quantity, is
significantly enhanced at later times in the counterrotating model S1414ct compared
to the irrotational (S1414) and corrotational (S1414co) model,
respectively. The same is true for, e.g., S1216ct compared to
S1216co. A direct consequence are differences in the strength of the postmerger
gravitational wave signal, which we calculate from the second
derivative of the quadrupole of the matter distribution using the
common quadrupole formula \citep[see ][]{oechslin2007}.

The NS spin setup also determines the total amount of angular momentum
in the system and thus directly influences the large-scale structure
of the merger remnant, the torus mass (Sect. \ref{sect:discmass}), and also the
amount of mass ejected from the system (Sect. \ref{sect:ejecta}). In corotating models, the total angular
momentum right after merging is larger by about 8\% compared to the irrotational case with the consequence that two
primary spiral arms form during merging, carrying this
additional angular momentum into a massive high-angular-momentum
torus. The central remnant is left
behind with a similar angular momentum content as in the irrotating
case (see the parameter $a_\mathrm{remnant}$ in Table
\ref{tab:inittable}). In the opposite, i.e. counterrotating case, the 8\% reduction of the
total angular momentum results in an even smaller torus and a more
compact remnant. A spin configuration of the ``tilted'' type as e.g. in model S1414t1 (large initial NS
spin in y-direction) leads to a remnant with a rotation axis parallel
to the total angular momentum direction (including NS spins) but
inclined relative to the orbital angular momentum of the initial
binary. It also results in a diffuse, puffed-up halo with no well-defined symmetry.

\subsubsection{Equation of state}
\label{sect:results-eos}

In this section, we investigate how the characteristic features of our EoSs
influence the merger dynamics and outcome. The supernuclear behaviour of the EoS determines the compactness of
the initial NSs (see Sect. \ref{sect:eos} and
Fig. \ref{fig:mrprofiles}).
In systems with larger NSs the plunge
and merging phase sets in earlier, i.e. at a larger orbital separation,
and thus, less energy and angular momentum is radiated away in such
models. For instance, in S1414 about 7\% of the total angular momentum
is lost while in LS1414 about 12\% and in A151414 about 15\% is
radiated away during inspiral. Thus, besides the spin setup, also the NS
compactness has a large influence on the angular momentum
retained in the system right after merging. The corresponding spin parameter
$a_\mathrm{system}$ is listed in Table \ref{tab:inittable}.

The supernuclear behaviour of the EoS also crucially affects the
structure of the merger remnant and its collapse time. For
instance, the remnants of the LS-EoS models both collapse to BHs
shortly after merger, either immediately (LS1216) or delayed (LS1414),
while in the corresponding Shen-EoS model S1414 and APR-EoS models
A151414 and A21414, no sign of a collapse can be seen within the
simulation time. The collapse time has a direct influence on the torus
mass, since secondary spiral arms, in which angular momentum transport
to matter at larger radii takes place, only form if the central
remnant does not collapse immediately.

The EoS stiffness in the supernuclear regime also affects the non-axisymmetry of the merger
remnant. As suggested by earlier simulations of bar-mode instabilities
in NSs, a
higher stiffness favours larger and longer-lasting non-axisymmetric
oscillations \citep[e.g.][]{houser1996,shibata2003barmode}. This mainly affects
the strength of the gravitational wave signal, but gravitational
torques transporting angular momentum to larger radii are also
amplified (see Sect. \ref{sect:discmass}).

The subnuclear EoS influences the merger and
postmerger dynamics only weakly. A comparison of models P1315 (ideal
gas EoS) and S1315 (Shen-EoS) yields similar remnant structures and
torus masses (see Fig. \ref{fig:discmass_vs_t}) but also similar
density profiles in the low-density regime outside the remnant where the
thermal contributions to the pressure become important. We cannot find
a clear reason for the latter observation. It might be caused by the fact
that the coefficient $\partial p/\partial
\epsilon |_{\rho}$ is much larger in the ideal-gas EoS at densities
below about $5\times10^{13}$ g/cm$^3$ (see
Fig. \ref{fig:p-rho-plot}). This might compensate the initially much
lower pressure in that density regime, once the energy density starts
to increase in that regime due to heating.

The additional thermal pressure provided by a higher value of
$\Gamma_\mathrm{th}$ in the APR-EoS leads to slightly ($\sim$ 10\%) lower central
densities and a less compact structure of the merger remnant in model
A21414 ($\Gamma_\mathrm{th}=2$) compared to model A151414 ($\Gamma_\mathrm{th}=1.5$). The same effect can
be observed as a result of the thermal pressure and energy in the Shen
case compared to the cold-Shen models. The additional thermal
pressure becomes more noticeable at lower densities. It leads to more extended (but less massive, see Sect. \ref{sect:discmass}) low-density halos around the merger remnants in models
A21414, S1315, and S1216 (``hot models'') compared to their analogues
A151414, C1315, and C1216 (``cold models''), respectively (also see
Sect. \ref{sect:thermal}).

\subsection{Thermal evolution of the finite temperature EoS-models}
\label{sect:thermal}

\begin{figure*}
\begin{center}
(a)\qquad
   \includegraphics[width=17cm]{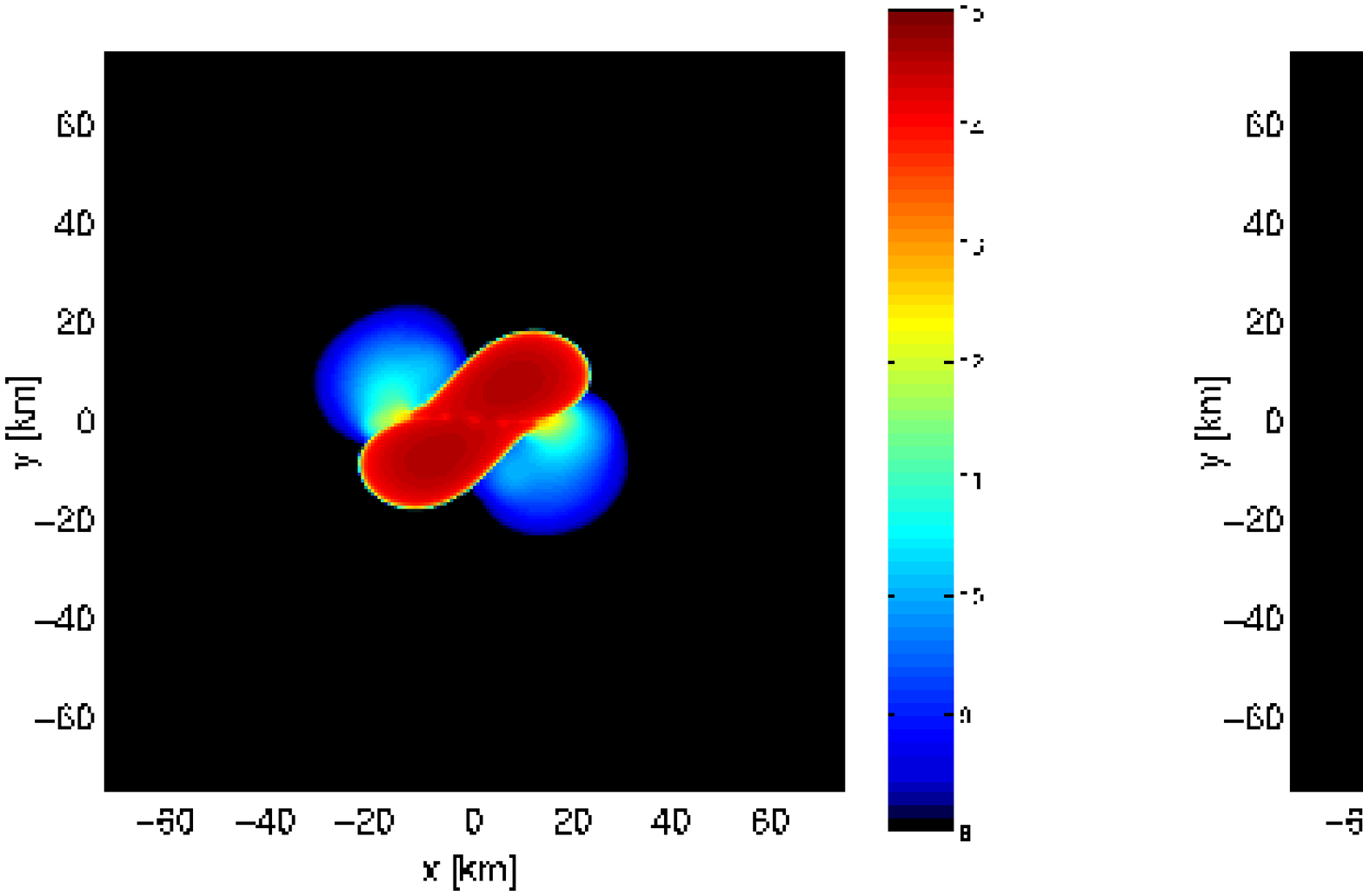}
\\
(b)
   \includegraphics[width=17cm]{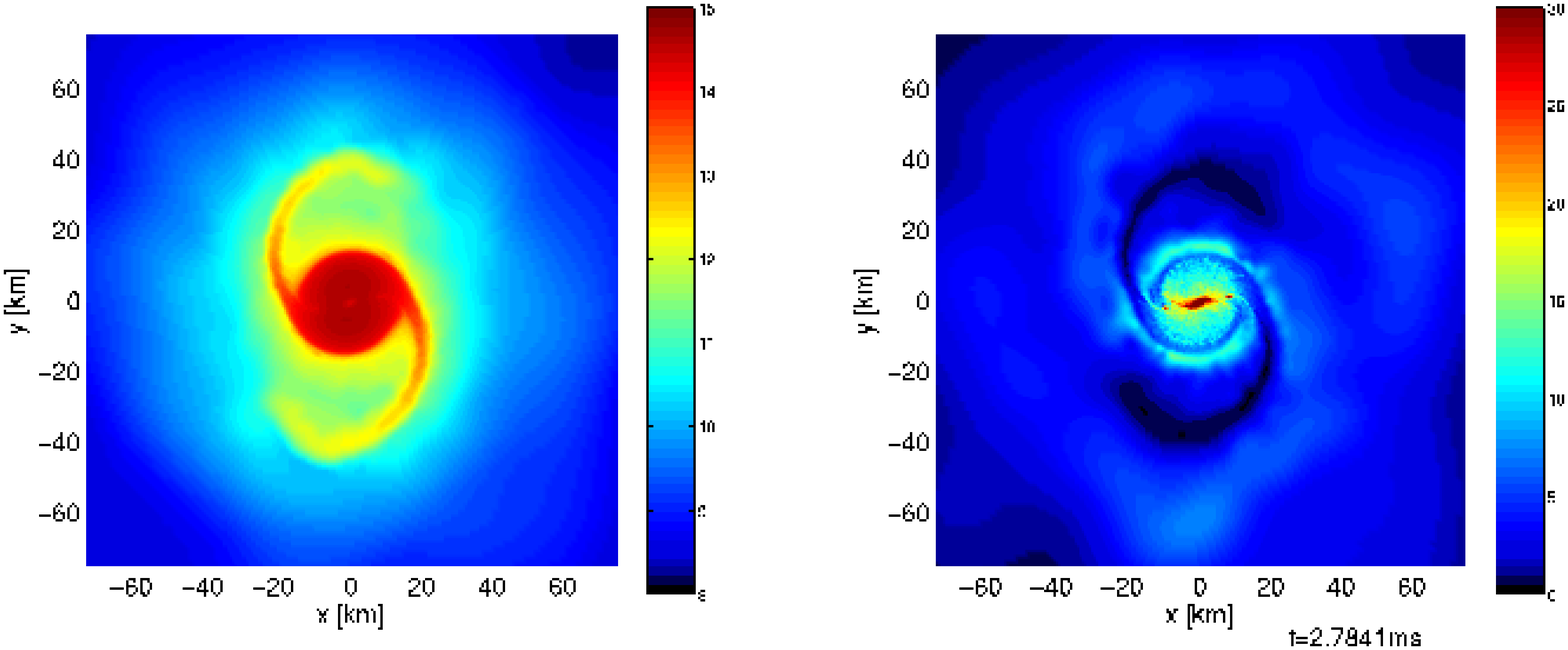}
\\
(c)
   \includegraphics[width=17cm]{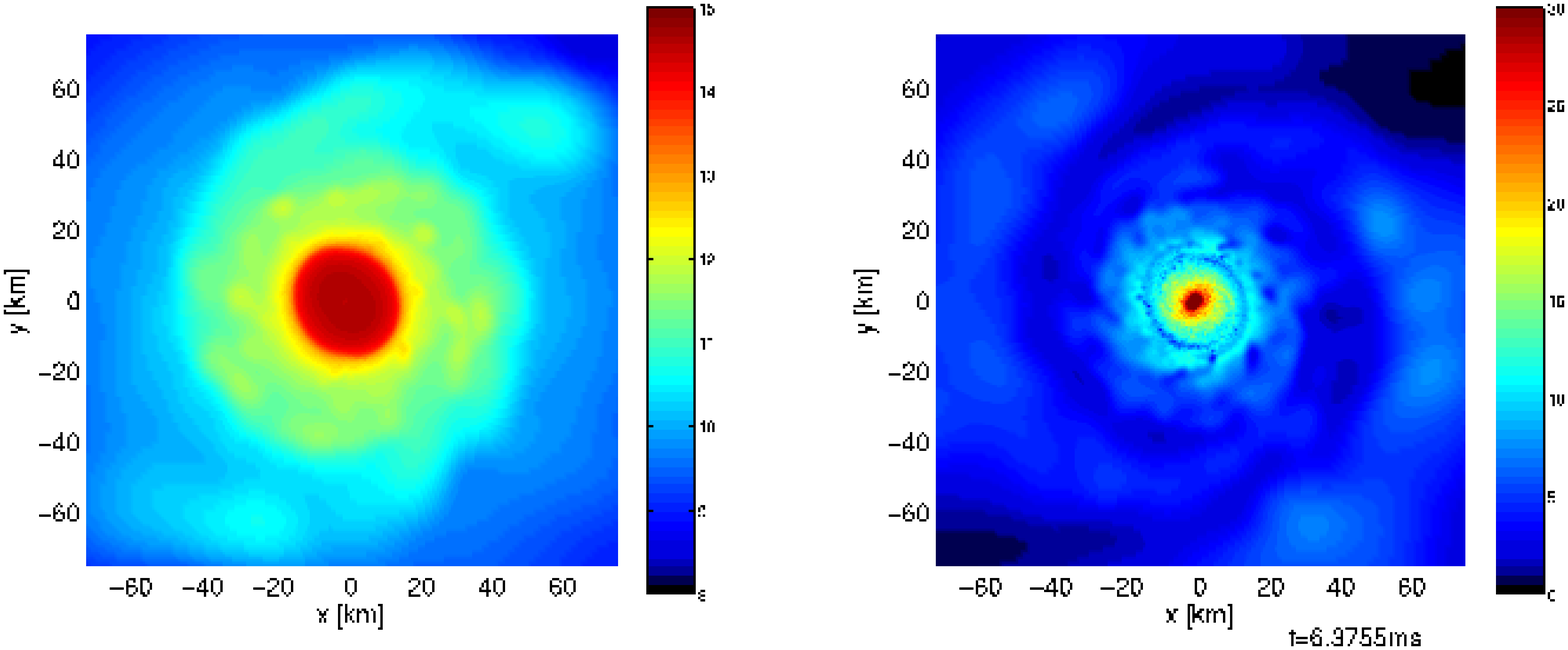}

  \caption{Density (left; logarithmic scale, in g/cm$^3$) and
  temperature (right; linear scale, in MeV) distribution in the orbital plane
  for model S1414 at the times given below the right panels. A primary spiral arm is missing but a pair of
  secondary spiral arms forms shortly after merging. Maximum
  temperatures are reached in the hot and KH-unstable vortex sheet
  between the two merging NSs (panel a, right). Visible are also the shock-heated layers at the surface of the merger remnant and along the
  leading edge of the spiral arms which plough through the less dense torus
  material (panel b, right). Panel (c) shows a later stage where the
  dense and hot merger remnant is surrounded by a ``warm'' torus.}
\label{fig:morph}
\end{center}
\end{figure*}

The initial NS models are set up with a temperature of 0.1 MeV, which is the lowest
value tabulated in the EoS tables. During inspiral, matter is heated
up in the center at most to a few MeV due to numerical viscosity, which however
has no dynamical consequences because thermal effects due to
temperatures of less than a few MeV yield a negligeable pressure contribution at the high densities in the
NS interior. There are three locations where significant heating
takes place during merging and the postmerger evolution: At the
Kelvin-Helmholtz(KH)-unstable collision
interface between the two NSs, at the remnant surface, and in the torus
itself (see Fig.\ref{fig:morph}).

The contact interface is
shock heated by the clash of the two NSs and by the turbulent motion
with viscous dissipation due to numerical viscosity in the
KH-vortices. Shocks also form at the surface of the non-axisymmetric, rapidly rotating merger remnant when it collides with the low-density
surrounding halo matter which has been ripped off the NS surfaces
along the shear interface at merging (see the temperature distribution
shown in Fig. \ref{fig:morph}, panels b and c and Fig. \ref{fig:morphxz}). Finally, heating takes place in the torus itself due to collision of
the dense spiral arms with each other or with ambient torus matter.

\begin{figure}
\begin{center}
   \includegraphics[width=8cm]{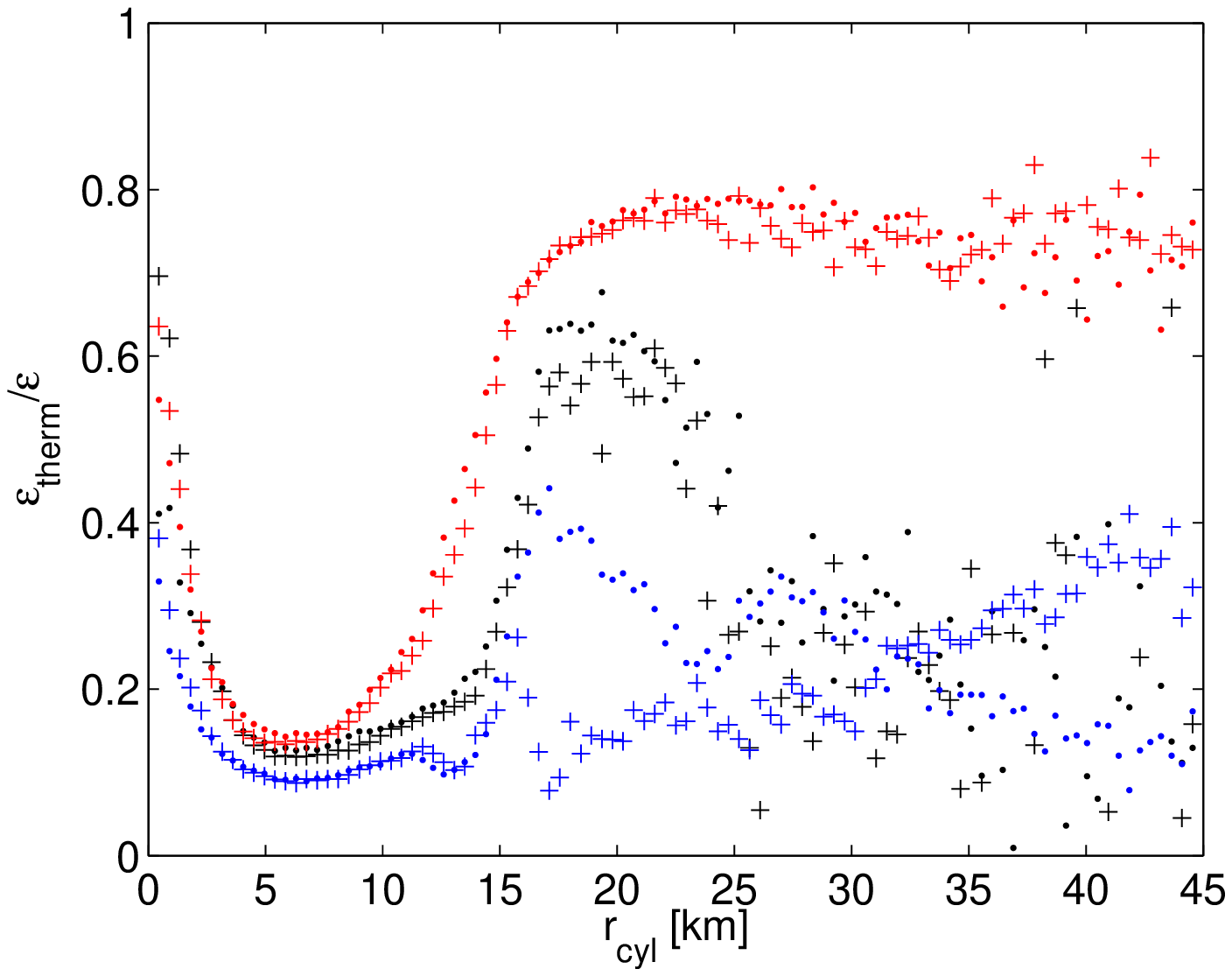}
   \includegraphics[width=8cm]{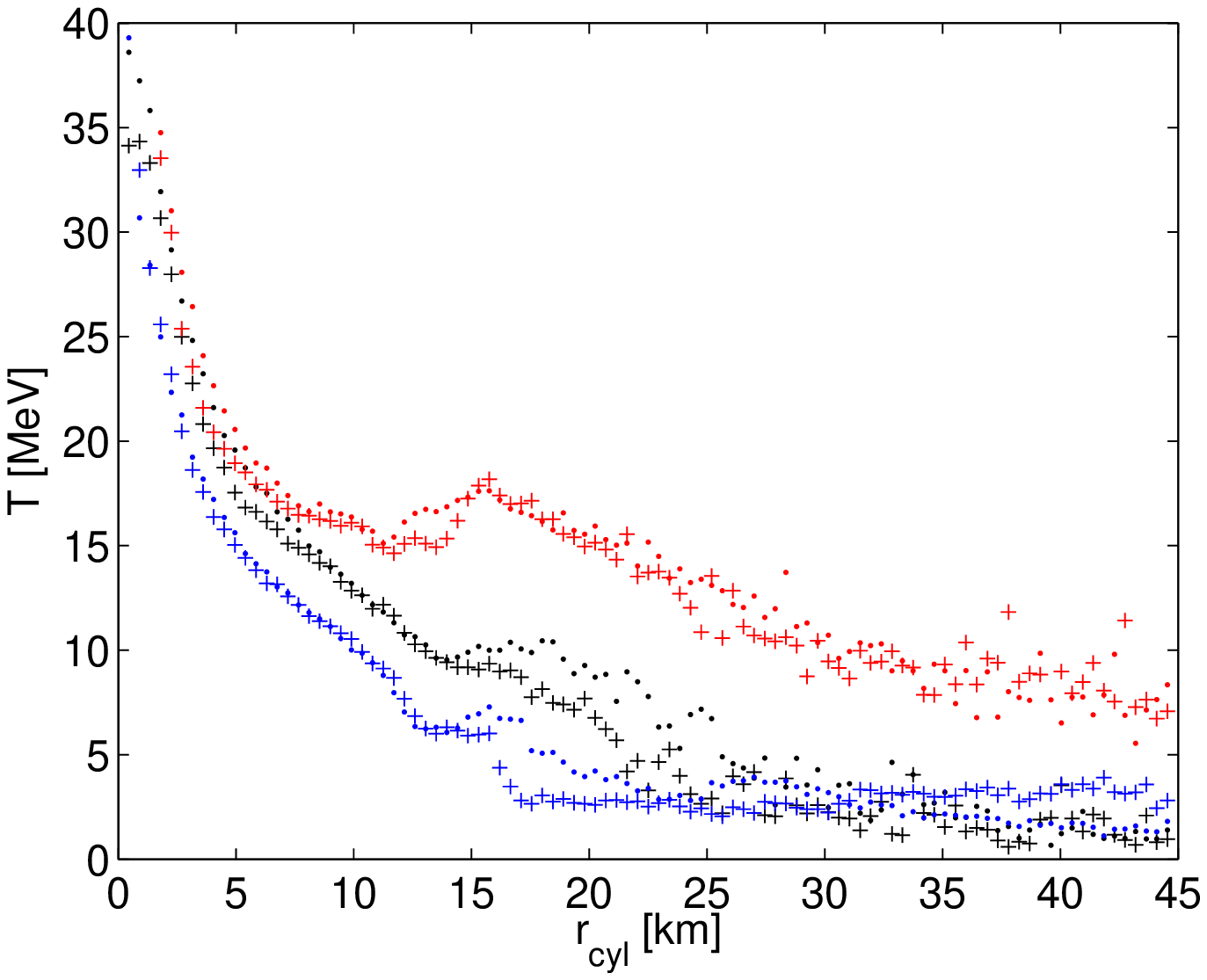}
\caption{Thermal energy contribution
$\epsilon_\mathrm{therm}/\epsilon=1-\epsilon(\rho,T=0,Y_{e})/\epsilon(\rho,T,Y_{e})$
relative to the total internal energy (top panel) and the corresponding temperatures as
given by the EoS (bottom panel) versus the cylindrical radius
$r_\mathrm{cyl}=\sqrt{x^2+y^2}$ at around 5ms (crosses) and
7ms (dots) after merging for three different models, S1414
(irrotating; in black), S1414co (corotating; in blue) and S1414ct
(counterrotating; in red). The plotted values are obtained by
angle-averaging the matter distribution in the orbital plane.}
\label{fig:epsilontherm}
\end{center}
\end{figure}

\begin{figure*}
\begin{center}

   \includegraphics[width=17cm]{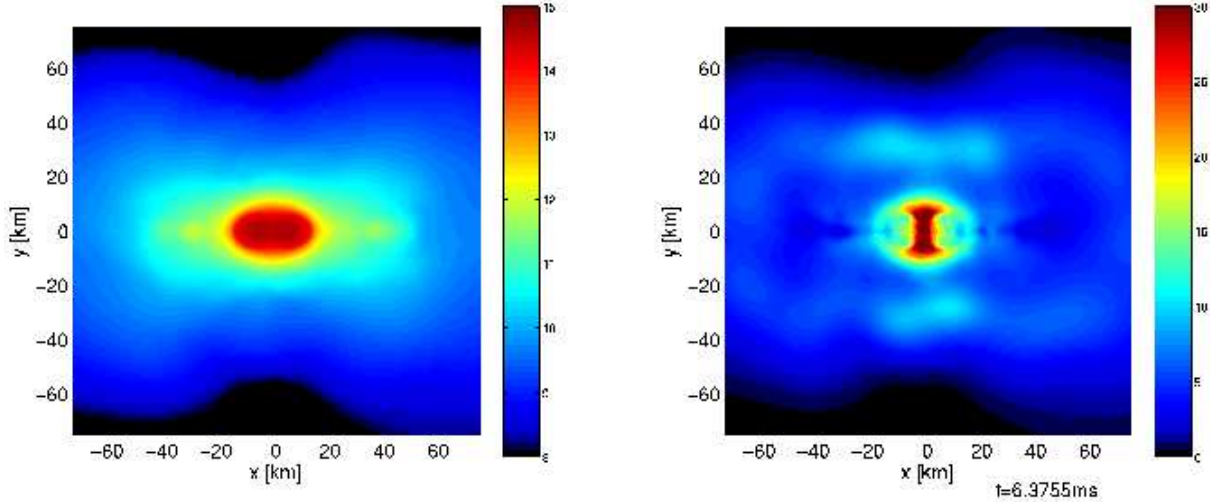}
\caption{Same situation as Fig. \ref{fig:morph}, lower panels, but perpendicular to the orbital
plane.}
\label{fig:morphxz}
\end{center}
\end{figure*}

\begin{figure*}
\begin{center}
   \includegraphics[width=16cm, height=5.3cm]{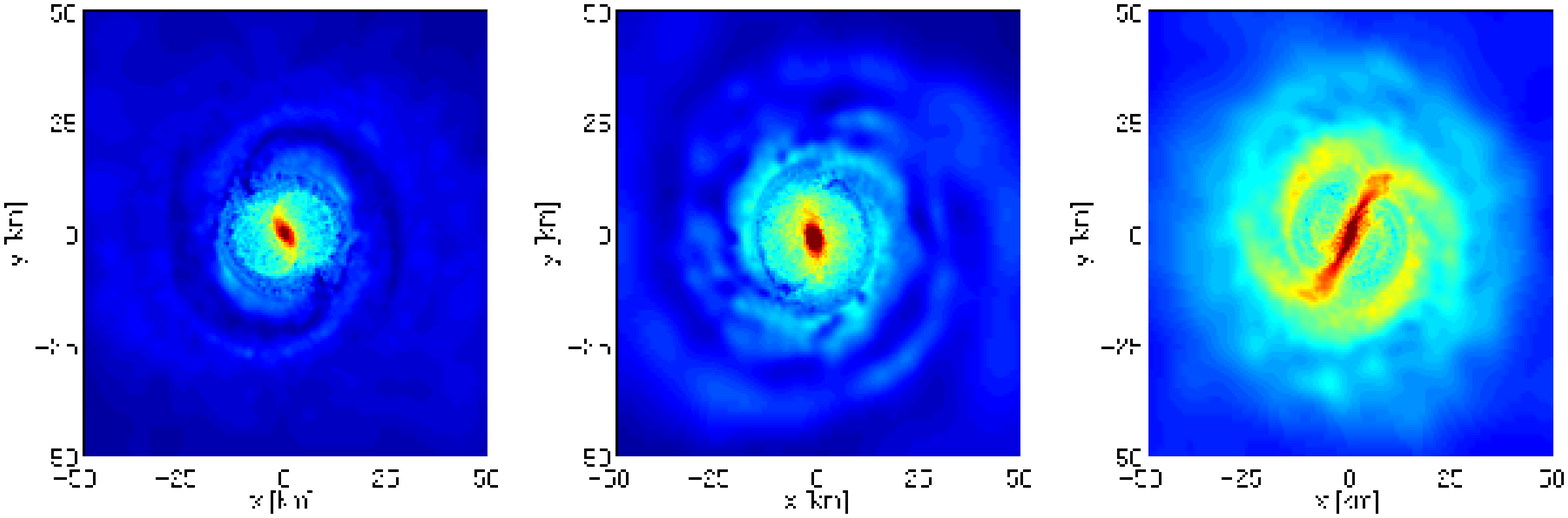}
	\includegraphics[width=0.9cm]{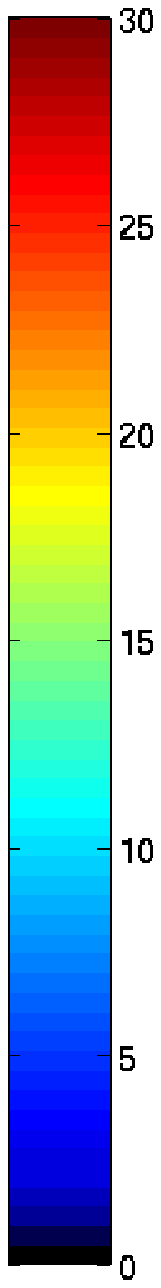}
\caption{Temperature distribution in the equatorial plane of the remnant and the torus shown
for models S1414co (corotating; left), S1414 (irrotating; middle) and S1414ct
(counterrotating; right) at $\sim 5$ms. Notice the different temperatures at
the collision interface of the two NSs and at the remnant surface.}
\label{fig:temp_vs_spin}
\end{center}
\end{figure*}

The strength of shock and viscous heating can be shown by considering
the thermal contribution
$\epsilon_\mathrm{therm}=\epsilon(\rho,T,Y_{e})-\epsilon(\rho,T=0,Y_{e})$ to
the internal energy $\epsilon$. In the upper panel of
Fig. \ref{fig:epsilontherm}, we show for the three symmetric models
S1414co, S1414 and S1414ct the ratio
$\epsilon_\mathrm{therm}/\epsilon(\rho,T,Y_{e})$, which is zero where no
shock heating is experienced and which tends to unity where the
degeneracy energy $\epsilon(\rho,T=0,Y_{e})$ is negligible compared to
the thermal component $\epsilon_\mathrm{therm}$. We clearly see that the
thermal contribution is largest in the counterrotating model and still
larger in the irrotating model than in the corotating case because
of the more extreme velocity jump at the contact interface. In
addition, the counterrotating model yields a larger non-axisymmetry of
the remnant and thus also leads to stronger shock heating at the remnant
surface (see Fig. \ref{fig:temp_vs_spin}). We also note that the torus
material in the counterrotating model S1414ct is hotter 
because it originates from the shock-heated
remnant surface rather than being ejected in unshocked spiral arms
as in models S1414co and S1414. In these models, the torus material is
initially cool but gets gradually heated up by collisions of the
spiral arms.

We generally obtain higher temperatures in the LS-EoS models than in
the Shen-EoS ones. These higher temperatures are accompanied by
the higher densities in the more compact merger remnants of the LS-models, which suggests that
the temperature difference is mainly caused by adiabatic compression
of the remnant interior.

\subsection{Torus mass estimates}
\label{sect:discmass}

\begin{figure}
\begin{center}
\includegraphics[width=7.5cm]{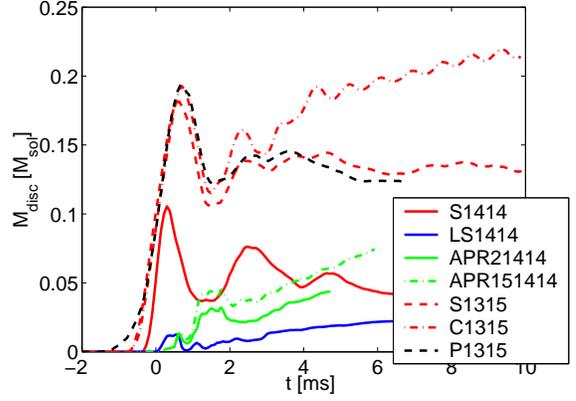}
\caption{The evolution of the torus mass for models with different
EoSs. Models with large initial NSs, as in case of the Shen-EoS
and the ideal-gas EoS (in red and black, respectively), show a rapid
rise of the torus mass already during the merging, while models with small initial NSs (in
blue and green, respectively) show
a slow but continuous increase of the torus mass setting in about 1ms after
merging. Models S1315 (red dashed) and P1315 (black dashed) illustrate
the similar torus mass evolution obtained with the Shen EoS and the
ideal-gas EoS. Finally, models A151414 and C1315 (to be compared
with A21414 and S1315, respectively) illustrate the
larger torus mass obtained using EoSs with reduced thermal pressure
contributions. Note: In contrast to \citet{oechslin2006a}, we have
included the non-compact $K_{ij}K^{ij}$ term in the central remnant
gravitational mass for calculating the torus mass here. Due to this, we
obtain slightly smaller values in the present work.}
\label{fig:discmass_vs_t}
\end{center}
\end{figure}

Prior to merging only very few fluid elements, e.g. in corotating models, have
a specific angular momentum which is large enough to fulfill the torus
criterion of Eq. (\ref{eqn:criterion}). But angular momentum is
transported from the central remnant outward to larger radii by
gravitational torques, which are generated by the forming
non-axisymmetric merger remnant during the merging and early postmerger
evolution. In models where a primary spiral arm is present, this transport
process is particularly efficient, and a rapid rise of the torus mass
occurs during the violent collision phase (see Fig. \ref{fig:discmass_vs_t} and Fig. \ref{fig:particles}, panel d). We find that a large part of the future torus matter
originates from the primary spiral arm. Secondary spiral
arms during the further evolution also contribute, although on a lower
efficiency level.

Test calculations with a physical setup similar to that of
model S1214 but with different particle numbers reveal that the torus mass is
not very sensitive w.r.t. the numerical resolution. The model using
$\sim 120000$ particles differs from the one using $\sim 280000$
particles by less than 5\% in torus mass.

With NS spins and EoS fixed, we find an increase of the torus mass
when the mass ratio drops below unity, while the total mass of the system
has only little influence \citep[see Fig. 2 in][]{oechslin2006a}. The
torus mass in case of the Shen EoS rises rapidly from about
0.04M$_{\odot}$ at $q=1$ to about 0.2M$_{\odot}$ at $q\simeq 0.8$ and
then roughly saturates for even smaller
$q$-values. For a $q$-value of 0.55 we find a torus mass of
about 0.25$M_{\odot}$.

\begin{figure}
\begin{center}
\includegraphics[width=7.5cm]{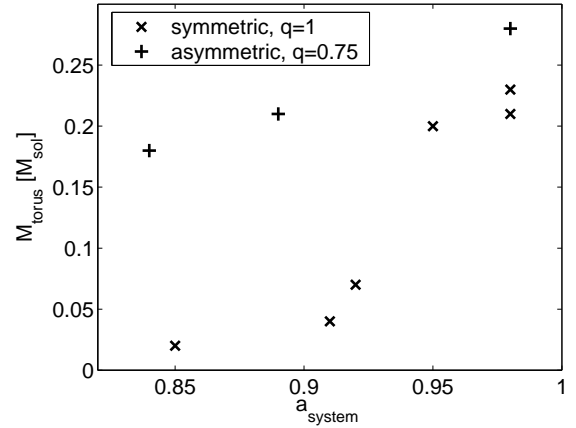}
\caption{Estimated torus masses versus total spin parameter
$a_\mathrm{system}$ of the systems measured immediately after merging. Plotted are two series of
models with mass ratio $q$=1 (crosses) and $q$=0.75 (plus signs),
respectively. In all models, the Shen-EoS was employed.}
\label{fig:discmass_vs_spins}
\end{center}
\end{figure}

When the EoS and the mass ratio are fixed, the torus mass increases
with the total amount of angular momentum available in the
system. Thus, corotating models yield a larger torus mass than
irrotating models, counterrotating smaller ones. Remarkably, models with a different
spin setup, but the same total angular momentum, yield similar
torus masses. This is the case for models S1414co (corotating) and
S1414t2 (tilted NS spins which sum up to a spin contribution aligned
with the orbital spin), and roughly also for models S1414 (no spins) and S1414o (oppositely oriented
spins). This suggests that the dependence of the torus mass on the NS spin setup can
be approximately reduced to a single parameter, the total angular momentum in the
system (see Fig. \ref{fig:discmass_vs_spins}).

Finally, the EoS influences the torus mass in various ways during the
different phases of the merging and postmerging evolution.
EoSs which yield less compact NSs like the Shen-EoS lead to lower
angular momentum loss during inspiral
(Sect. \ref{sect:results-eos}), thus favouring larger tori. Moreover, at
the time of plunging, more
angular momentum is transferred to the outermost NS matter sitting
opposite to the merger interface in case of larger NSs. This explains the
steep increase of the torus mass in the Shen-EoS models shortly after
merging (see Fig. \ref{fig:discmass_vs_t}). The corresponding models with
the same mass ratio and spin setup but using the LS-EoS and the APR-EoS
show a slow and continuous growth of the torus mass setting in only
about 1ms after merging.

During the postmerging evolution a second effect becomes important. The
more compact remnants allow for a more efficient angular momentum
transport through tidal torques from the central remnant to larger radii. This
effect is visible in case of model LS1414 and of the two APR-EoS
models. In the APR-cases it appears even more strongly due to the larger non-axisymmetry (see
Fig. \ref{fig:q22mode} panel b) of the remnant. A part of this growth, however, is likely to be caused by numerical
viscosity. This becomes clear at late times ($\geq 5$ms) in model LS1414, when
the remnant has already settled to a nearly axisymmetric state, but the
torus mass is still growing, although at a lower rate.
  
In models where the thermal pressure is neglected (C1216 and C1315) or
reduced (A151414), we observe larger torus masses compared to the
corresponding models S1216, S1315 and A21414, respectively. This
difference develops mainly during the early postmerger phase, which
suggests that the reason may be found in the different efficiency of
the angular momentum transport to the torus right after the actual merging
event. This claim is supported by the fact that in the cold models, the matter density is higher close to the surface of the newly formed remnant than in the hot models due to the lack of thermal
pressure. Thus, more matter is residing in the areas where
gravitational torques are strongest and more angular momentum and
ultimately mass is transferred outwards in the cold models.

\subsection{Estimated ejected mass}
\label{sect:ejecta}
We estimate the amount of ejected, i.e. gravitationally unbound, matter
by
considering for each fluid particle the quantity \citep{oechslin2002} 
$$
e_\mathrm{tot}=v^i\hu_{i}+\frac{\epsilon}{u^0}+\frac{1}{u^0}-1,
$$
which is conserved (in the comoving frame) if pressure forces are
small compared to gravitational forces and if
the metric is stationary. These conditions are well fulfilled for
cold gas in an axisymmetric, stationary torus, but are certainly only
approximately realized in our dynamical models. The quantity
$e_\mathrm{tot}$ asymptotes to the
Newtonian expression for the total energy $v^2/2+\epsilon-\phi$, $\phi$
being the Newtonian gravitational potential. For a particle to be
gravitationally unbound, this energy has to
be positive and we therefore use the criterion $e_\mathrm{tot}>0$ for
matter to be considered as ejected from the merger site.

\begin{figure*}
\begin{center}
  \begin{minipage}[t]{0.47\linewidth}
   \includegraphics[width=7.5cm]{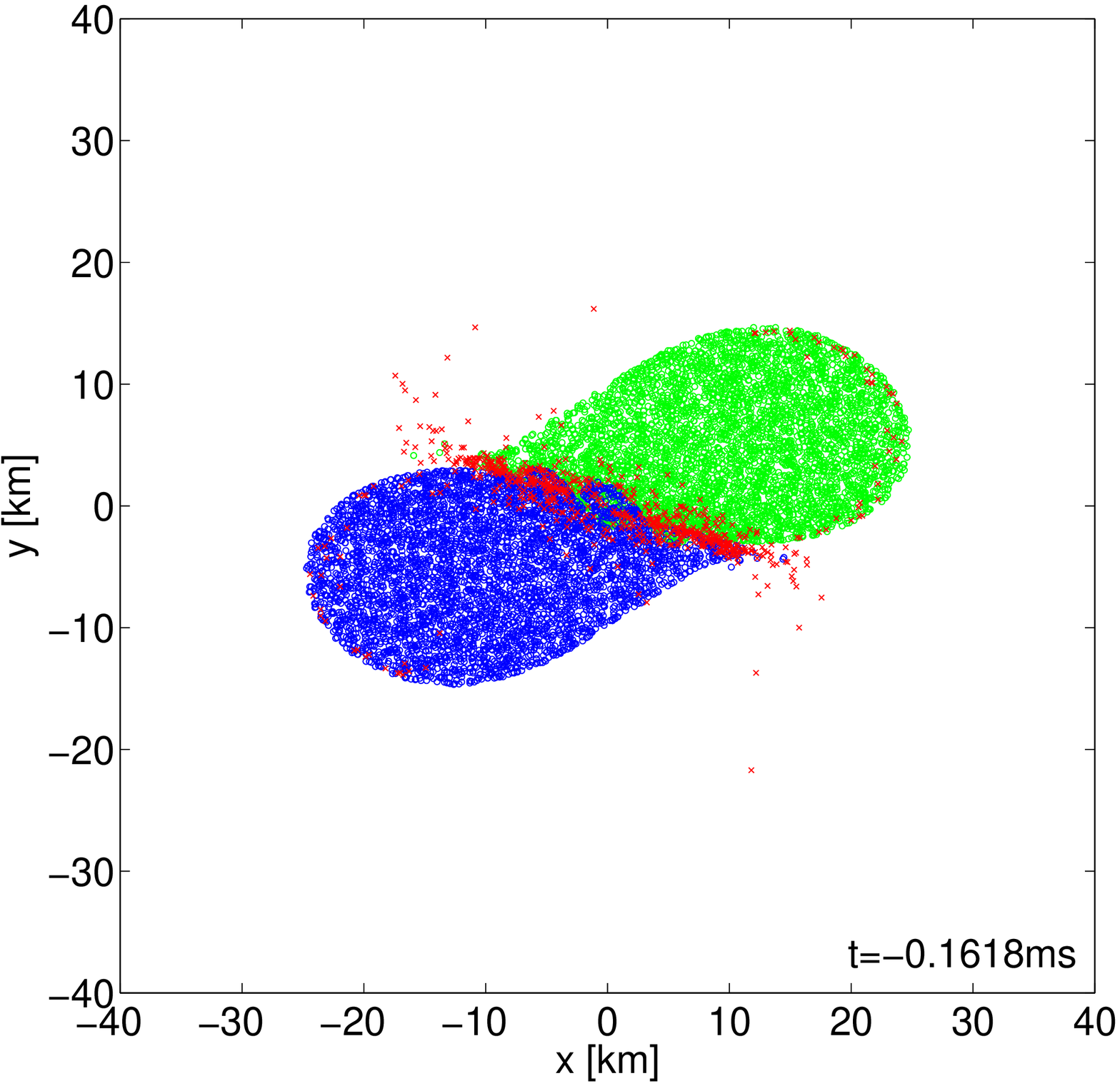}
\end{minipage}
  \begin{minipage}[t]{0.47\linewidth}
   \includegraphics[width=7.5cm]{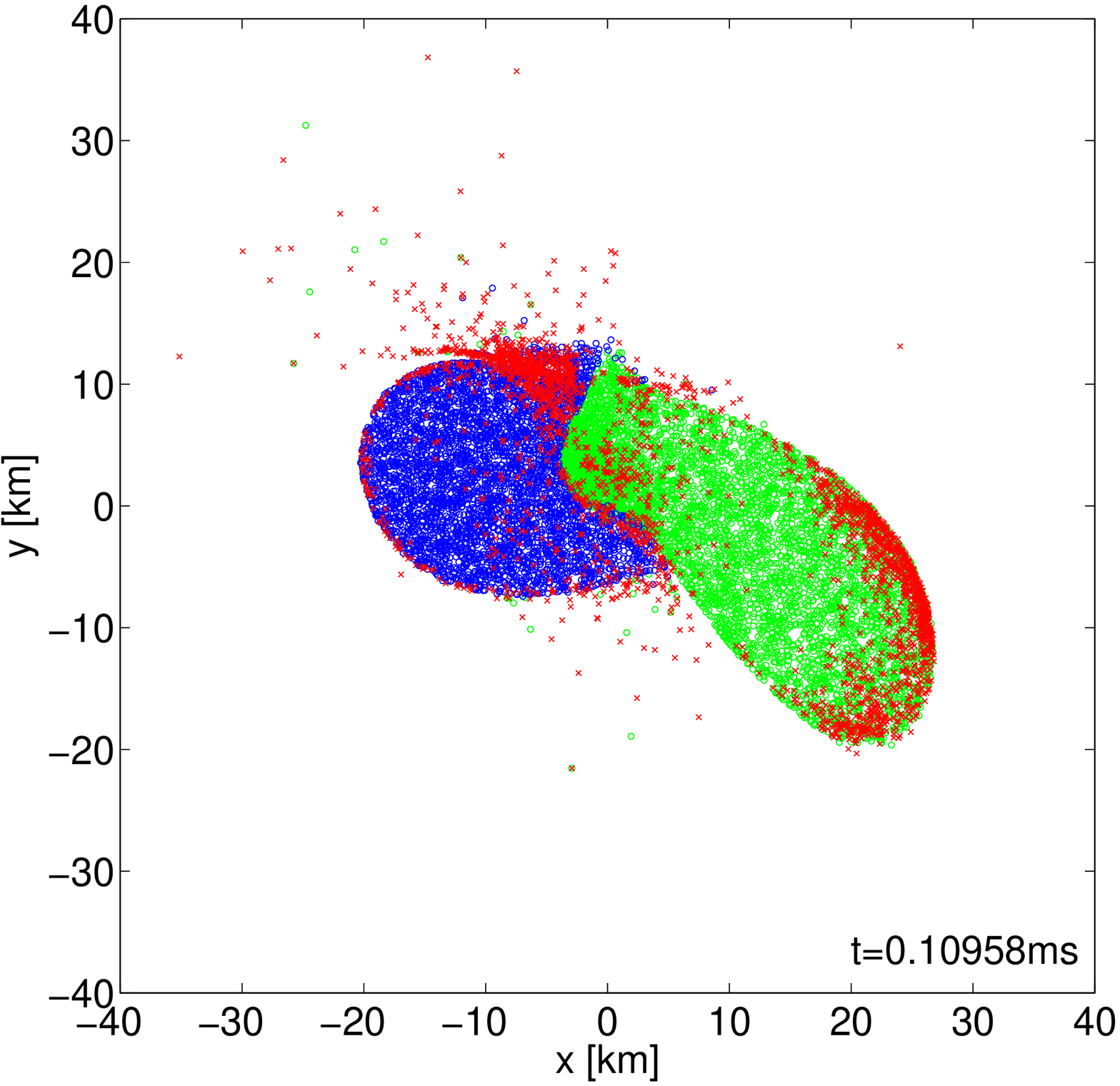}
\end{minipage}
\caption{Snapshots at the moment of merging of a symmetric (S1414; left) and an asymmetric
(S1216; right) model. The particles color-coded in red become gravitationally unbound
during the postmerger evolution. We can identify two sources of
ejected matter, the merger interface and the tip of the primary
spiral arm (if present). Note that the ejected matter is over-emphasized in
this plot since we plotted every second ejecta particle whereas for
the remaining matter only
every 10th particle near the equatorial plane is plotted. The red
particles at the contact interface are therefore squeezed out
perpendicular to the orbital plane.}
\label{fig:ejecta}
\end{center}
\end{figure*}

\begin{figure*}
\begin{center}
  \begin{minipage}[t]{0.46\linewidth}
   \includegraphics[height=7.5cm]{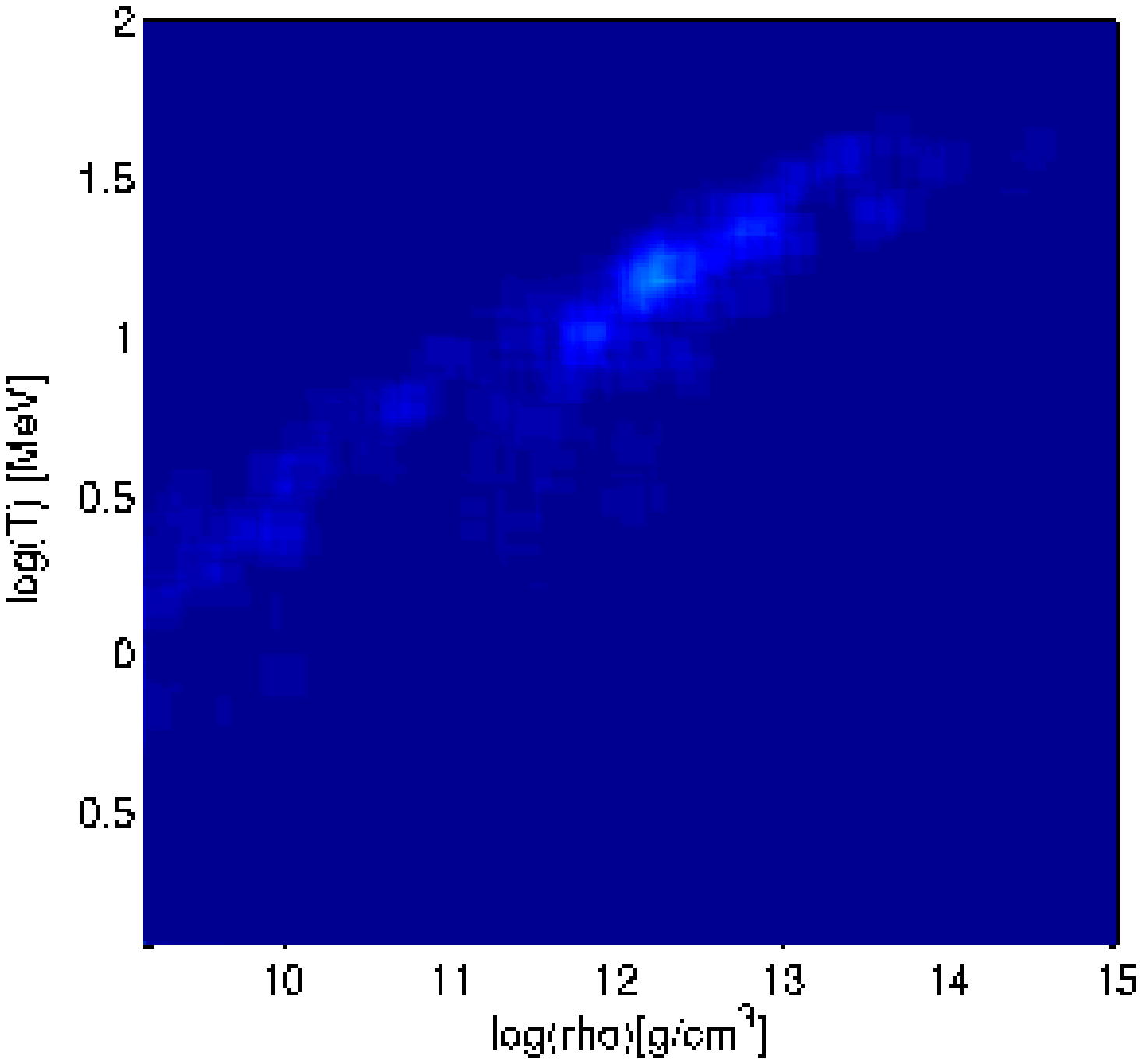}
\end{minipage}
  \begin{minipage}[t]{0.53\linewidth}
   \includegraphics[height=7.5cm]{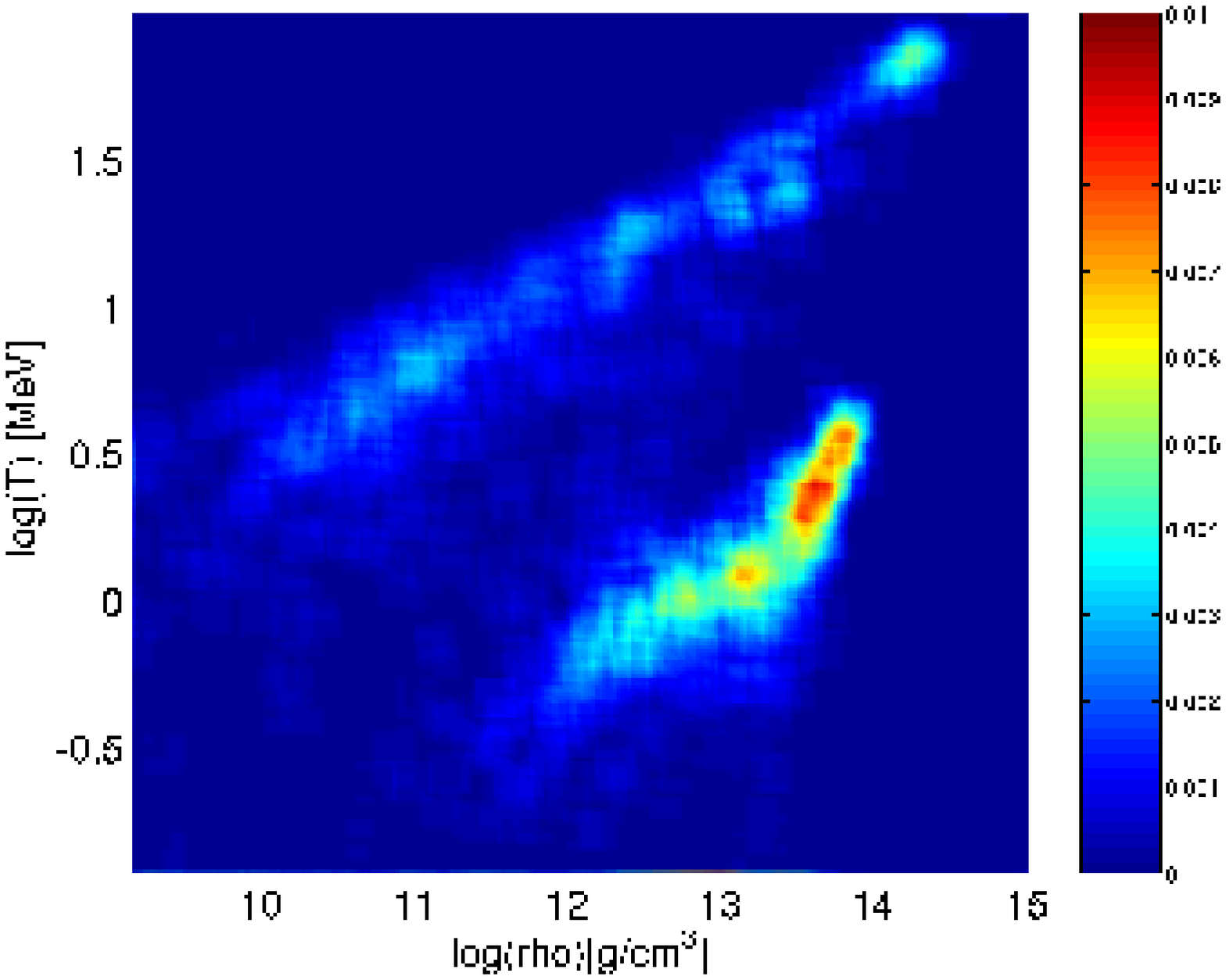}
\end{minipage}
\caption{Shown is the distribution of ejected matter in the
density-temperature plane around 1ms after merging for models S1414
(left) and S1216 (right). Color-coded is the amount of ejecta matter
per unit of density and temperature. We can identify two contributions to the ejecta, a
cold/low-entropy component and a hot/high-entropy one. The cold
component is only present in model S1216, which suggests that this
matter comes from the tip of the spiral arm, while the hot component
that exists in both cases stems from the merger interface.}
\label{fig:temperature_ejecta}
\end{center}
\end{figure*}

We find that ejected matter basically comes from two locations, the
tip of the spiral arms in models where such arms form, and the collision interface of the
merging NSs, where matter is
ejected perpendicular to the orbital plane (see
Fig. \ref{fig:ejecta}). Thus, we expect the ejected matter to consist
of a hot/high-entropy component from the shock-heated interface and of a cold/low-entropy component from
the tail of the primary spiral arm(s). In Fig. \ref{fig:temperature_ejecta}, we show
the distribution of the ejecta  for the two models S1414 and S1216 in the density-temperature
plane. We have color-coded the amount of ejecta mass per unit of density and
temperature. Clearly visible and well-separated are a
hot and a cold component. The latter is only present in the
asymmetric model S1216 where the cold component
originates from the tip of the spiral arm, while the hot component comes from
the shock-heated merger interface.

Generally, we find ejecta masses from about $10^{-3}$M$_\odot$ to
a few $10^{-2}$M$_\odot$ (Table \ref{tab:inittable}). The mass grows with the asymmetry of
the system. It also becomes larger, unlike the
torus mass, for counterrotating systems where the stronger shear flow
accelerates more hot matter at the collision interface to high velocities
without giving it a high angular momentum. Since the amount of ejected matter still increases at the end of our
simulations, we cannot provide precise numbers for the ejecta masses.

\subsection{Comparison with results of other works}

Model A21414 can be compared with model APR1414 of
\citet{shibata2006}. We obtain the same remnant object,
a hypermassive NS, and we find good agreement in the global
evolution, with somewhat higher values for the central lapse function,
i.e., smaller central densities in our model. The reason for this difference
might be the smaller angular momentum loss in GWs, which is slightly
underestimated by our backreaction scheme compared to the
simulation of \citet{shibata2006}. In our model $\sim 7\%$ of the initial angular momentum are radiated away during the last two ms
before merging, while it is $\sim 11\%$ in the calculation of \citet{shibata2006}. These
authors estimate a torus mass for their model of about
$0.02$M$_\odot$, which is slightly smaller than
our value (0.03 -- 0.04M$_\odot$). We expect the reason to be
again the smaller angular momentum loss by GWs in our model and
therefore the higher angular momentum of our postmerger configuration. It is
currently unclear to what degree this difference in torus masses could
also have numerical origin (e.g. SPH vs. grid-based calculation, resolution
issues, numerical viscosity) instead of being caused by the CFC
approximation. It is certainly desirable to attempt direct comparative
calculations with different codes and treatments of GR in the future. Considering an asymmetric model with the APR-EoS,
\citet{shibata2006} found a much lower sensitivity of the torus mass
to the mass ratio than we did for our models with the Shen-EoS. The reason
might be the much smaller NS radii in case of the APR-EoS. We have
seen in all our models with small NS radii (LS1414, LS1216, A151414,
A21414 and in a test case with a mass ratio of $q=0.875$ using the
APR-EoS) that the growth of the torus mass during merging is very small
compared to the growth setting in $\sim 1$ms after merging. Therefore, the dominant process for torus formation in
these cases is the angular momentum transport in the non-axisymmetric
postmerger remnant (see Sect. \ref{sect:discmass}). But at that stage,
mass-ratio effects are much smaller that at the stage of merging. This
might explain the much smaller mass ratio dependence observed by \citet{shibata2006} in their APR-models.

Our maximum temperatures at the center of the
remnants are generally considerably higher than those obtained for
similar models in Newtonian simulations \citep{ruffert2001,
rosswog2002a}, both with the Shen-EoS and the LS-EoS (e.g. S1414
compared to model C in \citealt{rosswog2002a} or model A64 in
\citealt{ruffert2001}). This can be explained by the stronger gravity in
our relativistic models. The consequently enhanced adiabatic compression
of the matter in the remnant leads to higher temperatures.

We also find a smaller torus mass in our model LS1414 compared to
model A64 in \citet{ruffert2001} which has been calculated with the
same binary parameters and the EoS but using Newtonian gravity. Our
calculations with the LS-EoS yield merger remnants which collapse BHs
within a few ms, while the Newtonian calculation yields a stable
merger remnant. 

\section{Conclusions}
\label{sect:conclusions}

We have carried out a set of relativistic binary NS merger calculations
starting from double NS binaries in quasiequilibrium, evolving them
through merging until the formation of a hypermassive NS surrounded by a thick torus.
We have considered two different physical, non-zero-temperature EoSs,
the Shen-EoS and the LS-EoS, an
ideal-gas EoS with parameters adapted to match the Shen-EoS at very
high densities and $T=0$, and
the zero-temperature APR-EoS with an ideal-gas-like extension to mimic
thermal effects as used
by \citet{shibata2006}. We have also considered different choices
for the initial NS masses and spins. The Einstein equations are solved
in the conformally flat approximation.

We have seen that the merger dynamics and outcome depend crucially not only
on the nuclear EoS, but also on the NS mass ratio and the NS
spins. The total mass in the system was shown to be less critical. Our results can be summarized in the following way:
\begin{itemize}
\item The angular momentum in the system is determined by the NS
spins but also by the size of the initial NSs. Corotation of the NSs
increases the total spin by about 8\% while counterrotation
reduces it by about the same amount. In systems with large
initial NSs, the final plunge and merging phase sets in earlier during
inspiral such that up to $\sim 8\%$ less angular momentum are radiated
away compared to models with small NSs. We find a clear correlation
between the angular momentum retained in the system and the torus mass.

\item The torus mass is very sensitive to the mass ratio, ranging from
about 1 -- 2\% of the (baryonic) mass in symmetric systems up to 
about 9\% of the system mass for a $q$-value of 0.55.

\item Merger remnants from asymmetric systems contain less angular
momentum and are thus more compact and more likely to collapse earlier to a
BH compared to their counterparts in symmetric systems.

\item The EoS determines not only the size of the initial NSs but also
influences the structure of the postmerger
remnant and its collapse time to a BH. Models using the stiff Shen-EoS
or APR-EoS do not show any sign of a collapse until the end of the
calculations at $\sim 10$ms after merging, while the soft LS-EoS leads
to an immediate (model LS1216) or delayed (model LS1414) collapse of
the remnant. The non-axisymmetry is largest in models using the
APR-EoS, whereas the remnant in the LS-EoS
model LS1414 quickly settles down to an axisymmetric state before
collapsing a few ms later. This mainly affects the strength of the
gravitational-wave signal, but angular momentum through gravitational
torques are also amplified. 

\item The influence of the thermal pressure contribution of the EoS has been investigated by
artificially setting the internal energy during the calculation to its
value at $T=0$. While the influence of $T\neq 0$-effects is rather small in the high-density regime
(10\% difference in the remnant central density), the thermal pressure
becomes very important in the low-density postmerger torus. If thermal
effects are neglected, postmerger tori are not only more compact but
also more massive. This torus-mass difference develops in the
early phase some ms after merging. The
reason may be found in a higher matter density close to the surface of
the newly formed compact remnant in the cold models due to the lack of thermal
pressure. Therefore, more matter resides in regions where
gravitational torques are strongest and more angular momentum and
ultimately mass is
transferred outwards in the cold models.

\item Around $10^{-3}-10^{-2}$M$_{\odot}$ become gravitationally unbound. Two
sources in the system, the merger interface and the tip of the primary spiral arm (only present in asymmetric systems), can be identified. Matter
from the merger interface undergoes shock heating and is therefore ejected
hot and entropy-rich, whereas matter from the spiral arm(s) remains cold. The ejected mass increases with the asymmetry of the system but
is also found to be sensitive to the spin state.  
 
\item Where direct comparison is possible, our results are
qualitatively similar to those obtained by \citet{shibata2006}. These authors
estimate a torus mass of $\sim 0.02M_\odot$ for their model APR1414
which is compatible with our value of $\sim 0.03-0.04M_\odot$ (still
growing at the end of the simulation) for our model A21414. Quantitative differences occur in the amount of
angular momentum carried away be GWs ($\sim 7\%$ in our model A21414 compared
to $\sim11\%$ in their model APR1414 during the last two ms before
merging) with implications for the compactness of the merger remnants.

\item We find generally higher temperatures and densities than those
obtained in Newtonian calculations \citep{ruffert2001,rosswog2002a}. 

\end{itemize}

Our relativistic DNS merger simulations have
extended previous work \citep[e.g.][]{oechslin2004, faber2004, shibata2005, shibata2006} to the investigation of models
with different NS masses, mass ratios, and spins,
using microphysical EoSs with a fully consistent
description of non-zero temperature effects in
comparison to simplified EoS treatments. Thermal
effects are irrelevant in the inspiral phase and for
determining the last evolutionary stages until the
binary becomes unstable. During the final plunge
and after the merging, however, the colliding stars
heat up and the structure and post-merging
evolution of the remnant depend on $T\neq 0$
effects, e.g. concerning the maximum density
and compactness and the possibility of mass and
angular momentum transfer from the remnant core
to a low-density torus. While thermal effects
tend to reduce the growth of such a torus
during the evolution of the merger remnant, they
significantly increase the mass ejection during
the merging because hot matter from the collision
interface of the two NSs can become gravitationally
unbound.

In a subsequent paper we will evaluate the
gravitational-wave signals from our mergers models,
characteristic features of which can provide
information about important EoS properties
\citep{oechslin2007}. More studies along two
lines also appear highly desirable. On the one hand,
a more detailed comparison between fully relativistic
models and the presented simulations with conformally
flat gravity is needed to assess the limitations of
the approximative treatment in the merger context. We
repeated some calculations found in another publication
and obtained qualitative agreement. The remaining
quantitative differences, however, could at least
partly be caused by resolution differences or
differences in the initial and boundary conditions,
or by differences in the numerical viscosity of the
schemes rather than the approach to GR. On the other
hand, many of the observed differences between the
models studied in this work could only be discussed
in a descriptive way. Many more calculations with
systematic variations of the initial conditions and
employed equations of state are needed to work out
dependences in detail and to develop a better
understanding of the causal links between model
input and results.

\begin{acknowledgements}

This work was supported by the Sonderforschungsbereich-Transregio 7
``Gravitational Wave Astronomy'' and by
the Sonderforschungsbereich 375 ``Astro-Particle Physics'' of the
Deutsche Forschungsgemeinschaft. The computations were performed at
the Rechenzentrum Garching and at the Max Planck Institute for Astrophysics.

\end{acknowledgements}

\appendix

\section{Hydrodynamics and field equations}
\label{app:hydro}

In the following, we briefly summarize the physical and numerical
framework we are using. We work with a 3+1 decomposition of spacetime, using
the following form of the metric $ds^2=g_{\mu\nu}dx^{\mu}dx^{\nu}$
\begin{equation}
ds^2=(-\alpha^2+\beta_i\beta^i)dt^2+2\beta_i dx^i dt+\gamma_{ij}dx^i
dx^j,
\end{equation}
where $\alpha$ is the lapse function, $\beta^i$ is the shift vector,
and $\gamma_{ij}$ is the spatial metric.\\
Imposing the conformally flat condition
$\gamma_{ij}=\psi^4\delta_{ij}$, where $\psi$ is the conformal factor, and choosing maximal slicing $\pa{t} K=K=0$, the Einstein equations can be written in
the following set of five elliptic equations for the metric elements
\citep[see e.g.][]{wilson1996,shibata1998}:
\begin{eqnarray}
\Delta(\alpha\psi)&=&2\pi\alpha\psi^5(\rho_E+2S)+\frac{7}{8}\alpha\psi^5 K_{ij}K^{ij}\label{eqn:lapse},\\
\Delta\psi&=&-2\pi\psi^5 \rho_E-\frac{1}{8}\psi^5 K_{ij}K^{ij}\equiv
4\pi\S_{\psi} \label{eqn:conffactor},\\
\Delta\beta^i&=&-\frac{1}{3}\pau{i}\pa{j}\beta^j\nonumber\\
&&+\pa{j}\ln\left(\frac{\alpha}{\psi^6}\right)\left(\pau{j}\beta^i+\pau{i}\beta^j
-\frac{2}{3}\delta^{ij}\pa{l}\beta^l\right)\nonumber\\
&&+16\pi\alpha \psi^4 j^i\label{eqn:shift}.
\end{eqnarray}
Here, $K_{ij}$ refers to the extrinsic curvature of the
$t=$const-3-slice and $K=K_i^i$ denotes its trace. In the
conformally-flat-maximal-slicing approximation, $K_{ij}$ can be
calculated directly from the metric elements
\begin{equation}
\label{kijform}
2\alpha\psi^{-4}K_{ij}=\delta_{il}\partial_j\beta^l+\delta_{jl}\partial_i\beta^l-\frac{2}{3}\delta_{ij}\partial_k\beta^k.
\end{equation}
Matter contributions enter the equations via the following
source terms
\begin{eqnarray}
\rho_E&=&n^\mu n^\nu T_{\mu\nu},\\
j^i&=&\gamma^i_\mu n_\nu T^{\mu\nu},\\
S_{ij}&=&\gamma_{i\mu} \gamma_{j\nu} T^{\mu\nu},
\end{eqnarray}
which refer to the matter energy density, the matter momentum density
and the spatial projection of the energy momentum tensor. Here,
$n^{\mu}$ denotes the unit normal vector to the $t=$const-3-slice.\\
Assuming a perfect fluid with an energy momentum tensor
\begin{equation}
T_{\mu\nu}=\rho h u_{\mu}u_{\nu}+pg_{\mu\nu},
\end{equation}
the first two matter contribution terms take the following form
\begin{eqnarray}
\rho_E&=&\rho h(\alpha u^0)^2-p,\\
j^i&=&\rho h \alpha u^0 u^\mu \gamma_\mu^i.
\end{eqnarray}
Here, $\rho$ is the baryonic rest mass density, $h=1+p/\rho+\epsilon$
the relativistic enthalpy, $p$ the fluid pressure, $\epsilon$ the
specific internal energy and $u^\mu$ the four velocity of the fluid.
The Lorentz factor $W=(1 + \gamma^{ij}u_i u_j)^{1/2}$ can be calculated using
the normalisation condition $u_{\mu}u^{\mu}=-1:$
\begin{equation}
\alpha u^0 = (1 + \gamma^{ij}u_i u_j)^{1/2}.\\
\end{equation}
The relativistic hydrodynamics equations are implemented in the following form
\begin{eqnarray}
\frac{d}{dt}\rst&=&-\rst\partial_i v^i\label{eqn:conteqn2},\\
\frac{d}{dt}\hu_i&=&-\frac{1}{\rho^*}\alpha\psi^6\partial_{i}
p-\alpha\hu^0\partial_{i}\alpha+\hu_j\partial_{i}\beta^j+\frac{2\hu_k\hu_k}{\psi^5\hu^0}\partial_{i}\psi\label{eqn:momentumeqn2},\\
\frac{d}{dt}\tau&=&-\frac{p}{\rho}\partial_i(v^i+\beta^i)\nonumber\\
&&-\frac{1}{\rst}(1-Wh)\pa{i} p (v^i+\beta^i)+S\label{eqn:energyeqn2},
\end{eqnarray}
where
\begin{eqnarray}
\frac{d}{dt}&=&\partial_t+v^i\partial_i,\\
\hu_i&=&hu_i,\\
\rst&=&\rho\alpha u^0\psi^6\label{eqn:rhoevol},\\
v^i&=&-\beta^i+\frac{\delta^{ij} u_j}{\psi^4 u^0}\label{eqn:vevol},\\
\tau&=&hW-\frac{p}{\rho W}-1-\frac{1}{2}\gamma^{ij}\hu_i\hu_j\label{eqn:eevol},\\
S&=&\frac{\hu_{i}\hu_j}{hu^0}K^{ij}(1-Wh)-\pa{i}\alpha(1-Wh)\psi^{-4}\hu_{i}\nonumber\\
&&-6\frac{p}{\rst}(v^i+\beta^i)\psi^5\pa{i}\psi.
\end{eqnarray}\\
While the continuity and momentum equations, Eqs. (\ref{eqn:conteqn2}) and
(\ref{eqn:momentumeqn2}), respectively, are commonly used in relativistic
hydrodynamics in this form, the energy equation (\ref{eqn:energyeqn2}) is
constructed out of the more commonly used variant 
\begin{eqnarray}
\label{eqn:energytotal}
\frac{d}{dt}\tau_0&=&-\frac{1}{\rst}\partial_i(\psi^6
p(v^i+\beta^i))\nonumber\\
&&+\frac{\hu_i\hu_j}{hu^0}K^{ij}-\frac{\hu_i}{\psi^4}\partial_i\alpha,
\end{eqnarray}
together with the momentum equation (\ref{eqn:momentumeqn2}) and the
Einstein equation to describe the time derivative of $\psi$ contained
in the spatial metric in expression (\ref{eqn:eevol}). Here, $\tau_0=hW-\frac{p}{\rho
W}-1$ is the total relativistic energy. In our simulations, Eq.
(\ref{eqn:energyeqn2}) yields more accurate
results than Eq. (\ref{eqn:energytotal}) in regions where
the kinetic energy dominates over the
internal energy part, e.g. at the surface of the moving NSs or
in cold regions of the torus. This is because $\tau$ is close to the internal energy
$\epsilon$, while $\tau_0$ contains contributions both from the
internal and the kinetic energies.

With the above set of hydrodynamics equations, we evolve the
quantities $\rst$, $\hu_i$ and $\tau$ forward in time using a 4th
order Runge-Kutta scheme. At each timestep, we have to recover the
primitive quantities $\rho$, $v^i$ and $\epsilon$ from the evolved
ones $\rst$, $u_i$ and $\tau$ using Eqs. (\ref{eqn:rhoevol}) -- (\ref{eqn:eevol}). These relations cannot be solved for the primitives in an analytical
closed form and have to be solved iteratively.\\
Since we do not implement any physics changing the electron fraction $Y_{e}$
in our fluid, the evolution of $Y_{e}$ is given by 
\begin{equation}
\frac{d}{dt}Y_{e}=0,
\label{eqn:yeeqn}
\end{equation}
i.e. $Y_{e}$ is simply advected with the fluid elements.

To integrate Eqs. (\ref{eqn:conteqn2}) -- (\ref{eqn:energyeqn2}) and
(\ref{eqn:yeeqn}), we use the Smoothed
particle hydrodynamics (SPH) method. Their implementation in SPH is fairly
straightforward as their form is similar to the Newtonian analogue, \citep[see e g.][]{siegler2000, oechslin2004}.

The Poisson-type metric equations (\ref{eqn:lapse}) -- (\ref{eqn:shift}) are solved on an overlaid
grid with a full multigrid solver. Since the unknowns also appear on the
RHS in the source terms, we have to iterate until convergence is
reached. Typically, if we start with the values from the previous
timestep as an initial guess, three or four iterations are enough. If we
start from a flat metric, i.e. $\alpha=\psi=1, \beta^i=0$ up to
10 iterations are necessary.\\
Boundary conditions and the extension beyond
the grid are obtained by a multipole expansion to quadrupole order of the formal solution 
$$
\Delta\psi=-4\pi S_\psi \Rightarrow \psi(\mathbf{r})=\int S_\psi(\mathbf{r'})/|\mathbf{r}-\mathbf{r'}|d^3\mathbf{r'}.
$$
The mapping of the necessary hydrodynamical quantities from the particles to the grid and the metric quantities back
to the particles is carried out with a third order interpolation.

\subsection{Artificial viscosity scheme}

In the hydrodynamics solver, we have implemented a new artificial viscosity (AV) scheme proposed by \citet{chow1997}, which is motivated by considering each
particle-neighbour pair as left and right states of a one-dimensional
Riemann problem. The quantities entering the resulting viscous
interaction between the two partners a and b are then the differences of the considered
physical quantities, projected along the line connecting the two
particles a and b and a signal velocity indicating the speed of a
signal sent from a to b, as seen in the computing frame.

Following \citet{chow1997}, we introduce an additional viscous
pressure term $\Pi_{ab}$ in the SPH momentum equation
\begin{equation}
\frac{d\hu_{i,a}}{dt}=-\alpha_a\psi_a^6\sum_b m_b\left(\frac{p_a}{{\rst_a}^2}+\frac{p_b}{{\rst_b}^2}+\Pi_{ab}\right)\pa{i}W_{ab}+R_1,
\end{equation}
where 
\begin{equation}
\Pi_{ab}=-\frac{Kv_{sig}}{\rst_{ab}}(\hu_{i,a}^*-\hu_{i,b}^*)e^i,
\end{equation}
when particles a and b are approaching. Otherwise, we set $\Pi_{ab}=0$.
Here, $K$ is a free parameter to adjust the AV strength, $v_{sig}$
stands for the signal velocity as discussed below,
$\mathbf{e}=\mathbf{r}_a-\mathbf{r}_b/|\mathbf{r}_a-\mathbf{r}_b|$ is the unit
vector from particle b to particle a and $\rst_{ab}=(\rst_a+\rst_b)/2$ is
the averaged coordinate conserved density. $W_{ab}$ is the SPH
interpolating kernel and the sum in the momentum equations runs over
all neighbouring particles. The remaining term $R_1$ stands for the
contribution arising due to the gravitational interaction. Note that
we have slightly adapted our SPH equations to general relativistic
hydrodynamics.

To guarantee that the viscous dissipation is positive definite, we
replace the usual $\hu_i$ by
\begin{equation}
\hu_i^*=h(v^i+\beta^i)\psi^4u^{0*},
\end{equation}
where
\begin{equation}
u^{0*}=W^*/\alpha=(1+\psi^{-4}(u_i e^i)^2)^{1/2}/\alpha
\end{equation}
is calculated with the Lorentz factor $W^*$ that involves the specific
momentum projected onto the line joining the two particles a and b.
To determine the viscous contribution to the change in $\tau$, we
first consider the SPH form of the total energy equation
(\ref{eqn:energytotal}):
\begin{eqnarray}
\frac{d\tau_0}{dt}&=&-\sum_b m_b \Bigg(\frac{\psi_a^6p_a (v^i+\beta^i)_a}{{\rst_a}^2}+\frac{\psi_b^6p_b
(v^i+\beta^i)_b}{{\rst_b}^2}\nonumber\\
&+&\Omega^i_{ab}\Bigg) \pa{i}W_{ab}+R_2,
\end{eqnarray}
where
\begin{equation}
\Omega^i_{ab}=-\frac{Kv_{sig}(\tau_{0,a}^*-\tau_{0,b}^*)e^i}{\rst_{ab}}
\end{equation}
if particles a and b are approaching, and $\Omega^i_{ab}=0$ otherwise. Here, $R_2$ stands for all additional terms due to gravity, and
$\tau^*_{0}$ is identical to $\tau_0$ with $W$ being replaced by $W^*$.\\
Using
\begin{equation}
\frac{d\tau}{dt}=\frac{d\tau_0}{dt}-\frac{\hu_i}{\psi^4}\frac{d\hu_i}{dt}+2\psi^{-5}\frac{d\psi}{dt},
\end{equation}
we finally end up, using also $\tau_0^*=\tau^*+(\hu_{i}e^i)^2/(2\psi^4)$, with the viscous contribution to the change in $\tau$:
\begin{eqnarray}
\frac{d\tau_a}{dt}&=&-\sum_b m_b\frac{Kv_{sig}}{\rst_{ab}}\Bigg(
(\tau^*_a-\tau^*_b)+\frac{(\hu_{i,a}e^i)^2}{2\psi_a^4}-\frac{(\hu_{i,b}e^i)^2}{2\psi_b^4}\nonumber\\
&-&\frac{\hu_{i,a}e^i(\hu_{j,a}-\hu_{j,b})e^j}{\psi_a^4}\Bigg)e^k\pa{k}W_{ab}+R_3,
\end{eqnarray}
where $R_3$ denotes the additional terms from hydrodynamical and
gravitational interactions.  
Since the term $\tau^*_a-\tau^*_b$ leads to a non-zero viscous contribution
even in the case $\mathbf{v}_a=\mathbf{v}_b$, we omit it in the final AV
implementation. Since this term is antisymmetric in a and b, this does
not lead to a violation of energy conservation.\\
Finally, the signal velocity $v_{sig}$ is approximated by
\begin{equation}
v_{sig}=c'_a+c'_b+|v^*_{ab}|,
\end{equation}
where $v^*_{ab}=(v_a^i-v_b^i)e^i$ is the relative speed projected
onto the line joining the two particles and 
\begin{equation}
c_a'=\frac{c_a+|v^*_{ab}|}{1+c_a|v^*_{ab}|}
\end{equation}
approximates the sound speed as seen in the local frame.

In addition, we implement a time-dependent viscosity parameter $K$ which
is increased in the presence of shocks, i.e. in regions where $\pa{i}
v^i$ is strongly negative and which decreases exponentially in regions with $\pa{i}v^i\gtrsim0$ to a constant minimal value \citep{morris1997}.

Comparisons of this viscosity scheme with the one used in the old code version
\citep{oechslin2004} show less particle noise and a smaller numerical
viscosity (see also similar work by \citealt{dolag2005} in a cosmological context).

\subsection{Radiation reaction and gravitational wave extraction}

The CFC approximation does not include the effects gravitational
radiation by construction. We therefore need a scheme to
mimic the GW backreaction onto the matter and a GW extraction
scheme. Following \citet{faye2003}, we implement the backreaction scheme
by adding a small, non-conformally flat contribution to the
metric. We implement their 3.5PN-accurate formalism, take, however, only into
account all 2.5PN-terms and the 3.5PN corrections related to the
gravitational potential and its derivatives. The reason is that during
the inspiral phase, where the backreaction is most important, PN
corrections are dominated by the internal gravitational potential
$U\simeq(\psi^4-1)/2\sim 0.3$ in the two NSs, whereas, e.g., velocity related
corrections are considerably smaller.

The following equations are solved
\begin{eqnarray}
\Delta U_5&=&-4\pi\sigma, \qquad \sigma=T^{00}+T^{ii},\\
\Delta R&=&-4\pi I_{ij}^{[3]}x^i\pa{j}\sigma,\\
\Delta U_7&=&-4\pi\rst(I_{ij}^{[3]}x^i\pa{j}U_5-R),\\
h_{00}&=&-\frac{4}{5}(1-2U_5)(I_{ij}^{[3]}x^i\pa{j}U_5-R)-\frac{8}{5}U_7,\\
h_{ij}&=&-\frac{4}{5}I_{ij}^{[3]},
\end{eqnarray} 
with \citep{BDS}
\begin{eqnarray}
I_{ij}&=&\int
   \rst\biggl\{x^ix^j\left(1+\frac{v^2}{2}-U+\epsilon\right)+\frac{11}{21}r^2v^iv^j-\nonumber\\
&&\frac{4}{7}x^ix^kv^jv^k+\frac{4}{21}v^2x^ix^j+\frac{11}{21}r^2x^i\pa{j}U-\nonumber\\
&&\frac{17}{21}x^ix^jx^k\pa{k}U\biggr\}.
\end{eqnarray}
Here, $r^2=x^ix^i$, $v^2=v^iv^i$, and $U$ is the Newtonian
gravitational potential. The first time derivative $I_{ij}^{[1]}$ of
the quadrupole is calculated
analytically using the above approximations and neglecting the total
time derivatives of $U, \pa{i}U, v^2, h$. These derivatives are
identically zero in the case of, e.g., a corotating binary. In the case
of non-corotating NS spins, a small error is introduced. The second and
third time derivatives, $I_{ij}^{[2]}$ and $I_{ij}^{[3]}$, respectively,
are calculated on-the-fly numerically with a finite difference approximation. 

\section{Definitions}
\label{sect:defs}

For further use, we define the following quantities.\\
Angular momentum:
$$
J=\int\rst\hu_{\phi}d^3x=\int\rst(x\hu_y-y\hu_x)d^3x,
$$
total baryonic mass:
$$
M_0=\int\rst d^3x,
$$
total gravitational mass:
$$
M=2\int S_{\psi}d^3x,
$$
and the orbital angular velocity:
$$
\Omega=\frac{\mathbf{r}_\mathrm{CM}\times\mathbf{v}_\mathrm{CM}}{r_\mathrm{CM}^2},
$$
where the index `CM' refers to the center of mass. 

\section{The torus criterion}
\label{app:disc}

We calculate the specific angular momentum $u_\phi$ of a test particle
rotating ($=\hu_\phi$, since $h=1$) on the innermost stable circular orbit (ISCO) around a BH as follows.\\
First, we determine the angular velocity $\Omega=d\phi/dt$ and specific angular momentum
 $u_{\phi}$ characterizing a given stationary circular orbit around a
BH with a fixed metric
$g_{\mu\nu}$ by solving the corresponding geodesic equation
$ \ddot x^r+\Gamma^r_{\mu\nu}\dot x^\mu \dot x^\nu =0 $. 
For a circular orbit in the equatorial plane, we have $\dot
r=\dot\theta=0$ and $\theta=\pi/2$ so that the geodesic equation reduces to
\begin{equation}
\label{eqn:omegaBH}
\Gamma^r_{tt}+2\Gamma^r_{\phi
t}\Omega+\Gamma^r_{\phi\phi}\Omega^2=0.
\end{equation}
Here, the $\Gamma^k_{ij}$ are the Christoffel symbols associated with
the BH metric. This equation specifies the angular velocity.\\
The specific angular momentum $u_{\phi}$ is then given by
\begin{equation}
\label{eqn:angmomBH}
u_{\phi}=g_{\phi\mu}u^{\mu}.
\end{equation}
We describe the BH resulting from the collapse of the merger remnant
by the following pseudo-Kerr metric \citep{grandclement2002} 
\begin{equation}
ds^2=-\alpha^2 dt^2+\psi^4(r^2\sin^2\theta(d\phi-N^\phi dt^2)^2+dr^2+r^2d\theta^2),
\end{equation}
where
\begin{eqnarray}
\alpha^2&=&1-\frac{2MR}{\Sigma}+\frac{4a^2M^2R^2\sin^2\theta}{\Sigma^2(R^2+a^2)+2a^2\Sigma
MR\sin^2\theta},\\
\psi^4&=&1+\frac{2M}{r}+\frac{3M^2+a^2\cos^2\theta}{2r^2}+\frac{(M^2-a^2)M}{2r^3}\nonumber\\
&&+\frac{(M^2-a^2)^2}{16r^4},\\
N^\phi&=&\frac{2aMR}{\Sigma(R^2+a^2)+2a^2MR\sin^2\theta},\\
R&=&r+\frac{M^2-a^2}{4r}+M,\\
\Sigma&=&R^2+a^2\cos^2\theta.
\end{eqnarray}
Here, $M$ and $a$ are the BH mass and spin parameter. These quantities
are identified with the gravitational mass and the spin parameter of
the compact merger
remnant. We define this object as the set of all fluid elements which are not found to
be in the torus.\\
Eqs. (\ref{eqn:omegaBH}) and (\ref{eqn:angmomBH}) then simplify to
\begin{eqnarray}
\Omega&=&\frac{-g_{\phi t,r}\pm\sqrt{g_{\phi
t,r}^2-g_{tt,r}g_{\phi\phi,r}}}{g_{\phi\phi,r}},\\
u_\phi&=&-\frac{g_{\phi\phi}\Omega+g_{\phi t}}{\sqrt{g_{tt}+g_{\phi
t}\Omega+2g_{\phi\phi}\Omega^2}},
\end{eqnarray}
with
\begin{eqnarray}
g_{tt,r}&=&-2\alpha'\alpha+2\psi^3rN^\phi(2rN^\phi+\psi'\psi
N^\phi\nonumber\\
&&+\psi r{N^\phi}'),\\
g_{\phi
t,r}&=&2\psi^3r(4rN^\phi\psi'+2\psi N^\phi+\psi r {N^\phi}'),\\
g_{\phi\phi,r}&=&2\psi^3r(2r\psi'+\psi).
\end{eqnarray}
Here, the prime stands for the derivative w.r.t. the radial coordinate $r$.\\
Now, $u_\phi$ can be minimized along the radial coordinate $r$ to find
the ISCO. This procedure has to be iterated a few times, since the
remnant mass and spin parameter themselves depend on the separation
between remnant and torus and thus on the ISCO.

\bibliographystyle{aa}	
\bibliography{../biblio}

\begin{thebibliography}{60}
\expandafter\ifx\csname natexlab\endcsname\relax\def\natexlab#1{#1}\fi

\bibitem[{Akmal {et~al.}(1998)Akmal, Pandharipande, \& Ravenhall}]{akmal1998}
Akmal, A., Pandharipande, V.~R., \& Ravenhall, D.~G. 1998, Phys. Rev. C, 58,
  1804

\bibitem[{Aloy {et~al.}(2005)Aloy, Janka, \& M{\"u}ller}]{aloy2005}
Aloy, M.~A., Janka, H.-T., \& M{\"u}ller, E. 2005, A\&A, 436, 273

\bibitem[{Ayal {et~al.}(2001)Ayal, Piran, Oechslin, Davies, \&
  Rosswog}]{ayal2001}
Ayal, S., Piran, T., Oechslin, R., Davies, M.~B., \& Rosswog, S. 2001, ApJ,
  550, 846

\bibitem[{Bardeen {et~al.}(1972)Bardeen, Press, \& Teukolsky}]{bardeen1972}
Bardeen, J.~M., Press, W.~H., \& Teukolsky, S.~A. 1972, ApJ, 178, 347

\bibitem[{Berger {et~al.}(2005)}]{berger2005}
Berger, E. {et~al.} 2005, subm. to Nature, astro-ph/0508115

\bibitem[{Bildsten \& Cutler(1992)}]{bildsten1992}
Bildsten, L. \& Cutler, C. 1992, ApJ, 400, 175

\bibitem[{Blanchet {et~al.}(1990)Blanchet, Damour, \& Sch{\"a}fer}]{BDS}
Blanchet, L., Damour, T., \& Sch{\"a}fer, G. 1990, MNRAS, 242, 289

\bibitem[{Blinnikov {et~al.}(1984)Blinnikov, Novikov, Perevodchikova, \&
  Polnarev}]{blinnikov1984}
Blinnikov, S.~I., Novikov, I.~D., Perevodchikova, T.~V., \& Polnarev, A.~G.
  1984, Soviet. Astron. Letters, 10, 177

\bibitem[{Bulik {et~al.}(2003)Bulik, Belczynski, \& Kalogera}]{bulik2003}
Bulik, T., Belczynski, K., \& Kalogera, V. 2003, in Gravitational Wave
  Detection, Proceedings of the SPIE, ed. M.~Cruise \& P.~Saulson, Vol. 4856,
  146

\bibitem[{Chow \& Monaghan(1997)}]{chow1997}
Chow, E. \& Monaghan, J.~J. 1997, Journal of Computational Physics, 134, 296

\bibitem[{Damour {et~al.}(2001)Damour, Iyer, \& Sathyaprakash}]{damour2001}
Damour, T., Iyer, B.~R., \& Sathyaprakash, B.~S. 2001, Phys. Rev. D, 63, 044023

\bibitem[{Damour {et~al.}(2002)Damour, Iyer, \& Sathyaprakash}]{damour2002}
Damour, T., Iyer, B.~R., \& Sathyaprakash, B.~S. 2002, Phys. Rev. D, 66, 027502

\bibitem[{Dolag {et~al.}(2005)Dolag, Vazza, Brunetti, \& Tormen}]{dolag2005}
Dolag, K., Vazza, F., Brunetti, G., \& Tormen, G. 2005, MNRAS, 364, 753

\bibitem[{Eichler {et~al.}(1989)Eichler, Livio, Piran, \&
  Schramm}]{eichler1989}
Eichler, D., Livio, M., Piran, T., \& Schramm, D.~N. 1989, Nature, 340, 126

\bibitem[{Faber {et~al.}(2004)Faber, Grandcl{\'e}ment, \& Rasio}]{faber2004}
Faber, J., Grandcl{\'e}ment, P., \& Rasio, F.~A. 2004, Phys. Rev. D, 69, 124036

\bibitem[{Faber \& Rasio(2000)}]{faber2000}
Faber, J. \& Rasio, F.~A. 2000, Phys. Rev. D, 62, 064012

\bibitem[{Faber \& Rasio(2002)}]{faber2002}
Faber, J. \& Rasio, F.~A. 2002, Phys. Rev. D, 65, 084042

\bibitem[{Faber {et~al.}(2001)Faber, Rasio, \& Manor}]{faber2001}
Faber, J., Rasio, F.~A., \& Manor, J.~B. 2001, Phys. Rev. D, 63, 044012

\bibitem[{Faye \& Sch{\"a}fer(2003)}]{faye2003}
Faye, G. \& Sch{\"a}fer, G. 2003, Phys. Rev. D, 68, 084001

\bibitem[{Fox {et~al.}(2005)}]{fox2005}
Fox, D.~B. {et~al.} 2005, Nature, 437, 845

\bibitem[{Gehrels {et~al.}(2005)}]{gehrels2005nature}
Gehrels, N. {et~al.} 2005, Nature, 437, 851

\bibitem[{Gourgoulhon {et~al.}(2001)Gourgoulhon, Grandcl{\'e}ment, Taniguchi,
  Marck, \& Bonazzola}]{gourgoulhon2001}
Gourgoulhon, E., Grandcl{\'e}ment, P., Taniguchi, K., Marck, J.-A., \&
  Bonazzola, S. 2001, Phys. Rev. D, 63, 064029

\bibitem[{Grandcl{\'e}ment {et~al.}(2002)Grandcl{\'e}ment, Gourgoulhon, \&
  Bonazzola}]{grandclement2002}
Grandcl{\'e}ment, P., Gourgoulhon, E., \& Bonazzola, S. 2002, Phys. Rev. D, 65,
  044021

\bibitem[{Houser \& Centrella(1996)}]{houser1996}
Houser, J.~L. \& Centrella, J.~M. 1996, Phys. Rev. D, 54, 7278

\bibitem[{Isenberg \& Nester(1980)}]{Isenberg1980}
Isenberg, J. \& Nester, J. 1980, in General Relativity and Gravitation, ed.
  A.~Held, Vol.~1 (New York: Plenum Press), 23

\bibitem[{Kochanek(1992)}]{kochanek1992}
Kochanek, C.~S. 1992, ApJ, 398, 234

\bibitem[{Lattimer {et~al.}(1985)Lattimer, Pethik, Ravenhall, \& Lamb}]{LLPR}
Lattimer, J.~M., Pethik, C.~J., Ravenhall, D.~G., \& Lamb, D.~Q. 1985, Nucl.
  Phys., A432, 646

\bibitem[{Lattimer \& Swesty(1991)}]{lattimer1991}
Lattimer, J.~M. \& Swesty, F.~D. 1991, Nucl. Phys., A535, 331

\bibitem[{Miller(2004)}]{miller2004}
Miller, M. 2004, Phys. Rev. D, 69, 124013

\bibitem[{Monaghan(1992)}]{monaghan1992}
Monaghan, J.~J. 1992, Annu. Rev. Astron. Astrophys., 30, 543

\bibitem[{Morris \& Monaghan(1997)}]{morris1997}
Morris, J.~P. \& Monaghan, J.~J. 1997, Journal of Computational Physics, 136,
  41

\bibitem[{Morrison {et~al.}(2004)Morrison, Baumgarte, \&
  Shapiro}]{morrison2004}
Morrison, I.~A., Baumgarte, T.~W., \& Shapiro, S.~L. 2004, ApJ, 610, 941

\bibitem[{Narayan {et~al.}(1992)Narayan, Paczy{\'n}ski, \& Piran}]{narayan1992}
Narayan, R., Paczy{\'n}ski, B., \& Piran, T. 1992, ApJL, 395, L83

\bibitem[{Nice {et~al.}(2005)Nice, Splaver, Stairs, Loehmer, Jessner, Kramer,
  \& Cordes}]{nice2005}
Nice, D.~J., Splaver, E.~M., Stairs, I.~H., {et~al.} 2005, ApJ, 634, 1242

\bibitem[{Oechslin \& Janka(2006)}]{oechslin2006a}
Oechslin, R. \& Janka, T. 2006, MNRAS, 368, 1489

\bibitem[{Oechslin \& Janka(2007)}]{oechslin2007}
Oechslin, R. \& Janka, T. 2007, in preparation

\bibitem[{Oechslin {et~al.}(2002)Oechslin, Rosswog, \&
  Thielemann}]{oechslin2002}
Oechslin, R., Rosswog, S., \& Thielemann, F.-K. 2002, Phys. Rev. D, 65, 103005

\bibitem[{Oechslin {et~al.}(2004)Oechslin, Ury\=u, Poghosyan, \&
  Thielemann}]{oechslin2004}
Oechslin, R., Ury\=u, K., Poghosyan, G., \& Thielemann, F.-K. 2004, MNRAS, 349,
  1469

\bibitem[{Oohara \& Nakamura(1989)}]{oohara1989}
Oohara, K. \& Nakamura, T. 1989, Prog. Theor. Phys., 82, 535

\bibitem[{Oohara \& Nakamura(1992)}]{oohara1992}
Oohara, K. \& Nakamura, T. 1992, Prog. Theor. Phys., 88, 307

\bibitem[{Paczy{\'n}ski(1986)}]{paczynski1986}
Paczy{\'n}ski, B. 1986, ApJL, 308, L43

\bibitem[{Paczy{\'n}ski(1991)}]{paczynski1991}
Paczy{\'n}ski, B. 1991, Acta Astron., 41, 257

\bibitem[{Rosswog \& Davies(2002)}]{rosswog2002a}
Rosswog, S. \& Davies, M.~B. 2002, MNRAS, 334, 481

\bibitem[{Rosswog {et~al.}(2000)Rosswog, Davies, Thielemann, \&
  Piran}]{rosswog2000}
Rosswog, S., Davies, M.~B., Thielemann, F.~K., \& Piran, T. 2000, A\&A, 360,
  171

\bibitem[{Rosswog \& Liebend{\"o}rfer(2003)}]{rosswog2003}
Rosswog, S. \& Liebend{\"o}rfer, M. 2003, MNRAS, 342, 673

\bibitem[{Rosswog {et~al.}(1999)Rosswog, Liebend{\"o}rfer, Thielemann, Davies,
  Benz, \& Piran}]{rosswog1999}
Rosswog, S., Liebend{\"o}rfer, M., Thielemann, F.~K., {et~al.} 1999, A\&A, 341,
  499

\bibitem[{Ruffert \& Janka(2001)}]{ruffert2001}
Ruffert, M. \& Janka, H.-T. 2001, A\&A, 380, 544

\bibitem[{Ruffert {et~al.}(1996)Ruffert, Janka, \& Sch{\"a}fer}]{ruffert1996}
Ruffert, M., Janka, H.-T., \& Sch{\"a}fer, G. 1996, A\&A, 311, 532

\bibitem[{Shen {et~al.}(1998{\natexlab{a}})Shen, Toki, Oyamatsu, \&
  Sumiyoshi}]{shen1998}
Shen, H., Toki, H., Oyamatsu, K., \& Sumiyoshi, K. 1998{\natexlab{a}}, Nucl.
  Phys. A, 637, 435

\bibitem[{Shen {et~al.}(1998{\natexlab{b}})Shen, Toki, Oyamatsu, \&
  Sumiyoshi}]{shen1998b}
Shen, H., Toki, H., Oyamatsu, K., \& Sumiyoshi, K. 1998{\natexlab{b}}, Prog.
  Theor. Phys., 100, 1013

\bibitem[{Shibata(1998)}]{shibatairrotational1998}
Shibata, M. 1998, Phys. Rev. D, 58, 024012

\bibitem[{Shibata {et~al.}(1998)Shibata, Baumgarte, \& Ury{\=u}}]{shibata1998}
Shibata, M., Baumgarte, T., \& Ury{\=u}, K. 1998, Phys. Rev. D, 58, 023002

\bibitem[{Shibata {et~al.}(2003)Shibata, Karino, \&
  Eriguchi}]{shibata2003barmode}
Shibata, M., Karino, S., \& Eriguchi, Y. 2003, MNRAS, 343, 619

\bibitem[{Shibata \& Taniguchi(2006)}]{shibata2006}
Shibata, M. \& Taniguchi, K. 2006, Phys. Rev. D, 73, 064027

\bibitem[{Shibata {et~al.}(2005)Shibata, Taniguchi, \& Ury{\=u}}]{shibata2005}
Shibata, M., Taniguchi, K., \& Ury{\=u}, K. 2005, Phys. Rev. D, 71, 084021

\bibitem[{Shibata \& Ury{\=u}(2002)}]{shibata2002}
Shibata, M. \& Ury{\=u}, K. 2002, Prog. Theor. Phys., 107, 265

\bibitem[{Siegler \& Riffert(2000)}]{siegler2000}
Siegler, S. \& Riffert, H. 2000, ApJ, 531, 1053

\bibitem[{Stairs(2004)}]{stairs2004}
Stairs, I.~H. 2004, Science, 304, 547

\bibitem[{Ury{\=u} \& Eriguchi(2000)}]{uryu2000}
Ury{\=u}, K. \& Eriguchi, Y. 2000, Phys. Rev. D, 61, 124023

\bibitem[{Wilson {et~al.}(1996)Wilson, Matthews, \& Marronetti}]{wilson1996}
Wilson, J.~R., Matthews, G.~J., \& Marronetti, P. 1996, Phys. Rev. D, 54, 1317

\end{thebibliography}
\end{document}